# Cross-Domain Radiomap Prediction for Multi-Scatterer Environment: A Spherical-Wave-Based Approach

Wenli Li, Bin Wang, *Member, IEEE*, Haiyan Fan, Yujie Zhang, *Member, IEEE*,
Guangxu Zhu, *Member, IEEE* and Yi Zhang

***Abstract* — Radiomap prediction has found extensive applications in network planning and optimization. The radiomap is implicitly determined by electromagnetic (EM) wave propagation. The evolution of wireless communication toward higher frequency spectra has highlighted the effect of mesoscopic (i.e. wavelength-comparable scale) scatterers, which is negligible in the sub-6 GHz spectrum. To address this challenge, we develop a linear forward channel model to capture the propagation behavior in the multi-scatterer environment. The proposed model utilizes spherical-wave mode expansion to track the source radiation pattern, including scattering from a single scatterer and interactions among multiple scatterers. Both phenomena are represented as a superposition of spherical-wave modes, effectively capturing the multipath effect from a wave-based perspective. This forward model is then employed to formulate an inverse optimization problem, where the scattering responses and scatterer locations are jointly learned from sparse field measurements. Simulation results demonstrate that the proposed model accurately reconstructs and extrapolates the radiomap in both the spatial and the beam domain, further enables frequency-domain channel interpolation.**

## I. INTRODUCTION

ACCURATE and localized site-specific channel modeling is becoming increasingly indispensable for future wireless communication systems. As wireless networks evolve toward millimeter-wave, sub-terahertz, extremely large-scale antenna arrays, and integrated sensing and communication (ISAC), the channel response becomes increasingly sensitive to its surrounding environment, user location, beam direction, blockage state, and local scattering structures [1]-[5]. In such systems, the wireless channel is no longer merely a statistical abstraction of large-scale path loss and clustered multipath components, but rather a fine-grained electromagnetic response determined by the intricate interactions between radio waves and the physical environment. Therefore, reliable localized channel models are essential for channel state information (CSI) prediction, beam management, link adaptation, network planning, digital-twin-based evaluation, and reliable offline optimization of wireless system configurations prior to actual deployment [6]-[11].

Existing standardized and geometry-based channel models have provided efficient and widely adopted tools for wireless system evaluation [1], [12], [13]. Geometry-based stochastic models (GBSMs), for example, represent multipath propagation through statistically generated clusters, delays, angles, and Doppler components, efficiently reproducing important channel statistics for system-level simulations [12], [13]. However, since the scattering clusters and propagation parameters are mainly generated from statistical distributions, such models cannot be used for reconstructing fine-grained, site-specific channel responses in a given local environment. This limitation becomes more pronounced in high-frequency systems, where small variations in receiver position, beam direction, or nearby objects may lead to significant changes in dominant paths, received power distribution, and beam selection [6], [7].

Deterministic propagation models, especially ray-tracing-based methods, provide a more site-specific alternative by exploiting the geometric information of the propagation environment [14]-[17]. By simulating reflection, diffraction, scattering, and transmission paths, ray tracing can simulate dominant propagation in complex indoor and outdoor scenarios. Nevertheless, the accuracy of ray tracing strongly depends on high-fidelity three-dimensional maps, object geometries, the electromagnetic properties of materials, which are often expensive or difficult to obtain in practical deployments. Moreover, conventional ray-based and high-frequency asymptotic methods are effective only when the involved objects are electrically large and the propagation mechanisms can be well approximated by geometrical optics, the uniform theory of diffraction, or physical optics [18]. When environmental scatterers have characteristic sizes comparable to the wavelength, diffraction, coherent scattering, and multi-scatterer coupling may noticeably alter the local channel response, rendering ray-based approximations less reliable.

Full-wave computational electromagnetic methods, such as the finite-difference time-domain method, finite element method, and method of moments, can directly solve Maxwell's equations and accurately capture wave effects, including near-field interactions and multiple scattering [19]-[22]. However, their computational and storage costs are often prohibitive for

This work is supported by Guangdong Major Project of Basic and Applied Basic Research (No. 2023B0303000001), Zhixin Program of Shenzhen Research Institute of Big Data (No. J00220250003) and the NSFC under Grant 52501418. (*Corresponding author*: *Yi Zhang*)

Wenli Li, Bin Wang, Guangxu Zhu and Yi Zhang are with Shenzhen Research Institute of Big Data and Chinese University of Hong Kong (Shenzhen), Shenzhen 518172, China (e-mail: zhangyi@sribd.cn).

Haiyan Fan is with School of Mechanical Engineering, Southeast University, Nanjing 211189, China.

Yujie Zhang is with the School of Electrical and Electronic Engineering, Nanyang Technological University, Singapore 639798.

frequent channel reconstruction, large-scale receiver-grid prediction, or learning-based calibration in wireless communication scenarios. This creates a fundamental tradeoff in site-specific channel modeling: stochastic models are efficient yet lack sufficient environmental awareness; ray-based methods are geometry-aware but rely on detailed maps and high-frequency approximations; and full-wave methods are physically accurate but computationally intractable for system-level tasks. A practical channel modeling framework should reach a balance between physical consistency, computational efficiency as well as learnability.

Recently, learning-based channel modeling has attracted increasing attention in site-specific wireless channel reconstruction. Neural radio-frequency radiance fields, Gaussian-splatting-based radio field representations, and channel knowledge maps can learn continuous or compact representations of local channel responses from measurement data [8], [23]-[26]. Furthermore, generative artificial intelligence, particularly diffusion models, has recently been leveraged to construct dynamic and three-dimensional radio maps conditioned on environmental layouts [27]-[29]. By learning the complex distribution of propagation data, these generative approaches are capable of predicting high-fidelity spatial signal variations without requiring dense sampling. Overall, these methods provide new opportunities for reconstructing spatially varying channel information, predicting channel responses at unmeasured locations. However, methods of this nature rely on data-driven function fitting, where the underlying electromagnetic propagation mechanisms are only implicitly encoded. As a result, these methods cannot maintain a reliable performance if there exists a mismatch between the training data and the realistic environment.

This motivates physics-embedded learning for site-specific channel modeling. Instead of treating the channel as a black-box function, physics-embedded learning incorporates electromagnetic propagation regularity into the model structure and exploits wireless measurements to calibrate the remaining unknown parameters [30], [31]. Such a paradigm not only reduces the dependence on heavy training data but also improves generalization and interpretability. In particular, when the local environment contains medium-scale scatterers whose size matches the wavelength, explicit modeling of scattering and inter-scatterer coupling becomes indispensable. These scatterers may induce diffraction, resonant scattering, coherent interference, and multiple scattering interactions, which are difficult to capture using solely stochastic model, ray tracking or neural networks.

ISAC provides a new opportunity to bridge environmental perception with channel modeling. In ISAC-enabled wireless systems, dedicated waveforms or pilots are embedded to base stations to extract target-related parameters, such as range, angle, Doppler, and echo intensity [32]-[35]. By associating these measurements across antennas, time slots, and frequency bands, the surrounding environment can be characterized by sensing-derived point clouds that describe significant scatterers. Point clouds of this nature provide informative geometric priors for site-specific channel modeling, while CSI measurements collected from communication links can be used to monitor channel state variations in the receiving grid, frequency, and beam domains. By judiciously utilizing this information in a joint paradigm, channel modeling can be improved in terms of both physical consistency and learnability.

In this paper, we propose a physics-embedded supervised learning framework for site-specific wireless channel modeling. The proposed framework represents the localized channel response through a modal electromagnetic multi-scatterer formulation. Specifically, scalar spherical-wave expansions are used to describe spatial propagation modes, scatterer responses are represented by T-matrices, and source–scatterer–receiver interactions are connected through translation operations [36]-[38]. In this formulation, the wireless channel is expressed as a finite-dimensional coupled system, where the positions and scattering responses of local scatterers serve as trainable latent parameters. ISAC-induced point clouds are used to initialize or constrain scatterer positions, while CSI samples measured under different receiver locations and beam patterns are used as supervision so as to optimize the physical channel representation.

Compared to learning-based methods that are black-box in nature, the proposed framework embeds explicit electromagnetic scattering mechanisms into the learning process to guarantee physical consistency. In addition, our method avoids discretizing the entire region of interest (ROI) and forms a complexity-reduced representation of environment, thereby alleviating the inherent challenges in full-wave-based methods. At last, unlike ray-based methods, the proposed method can capture multi-scatterer coupling and wavelength-comparable scattering effects through a controllable modal formulation. In summary, our proposed method provides a physically interpretable, measurement-trainable, and complexity-controllable approach for reconstructing site-specific wireless channels.

The main contributions of this paper are summarized as follows:

1) We propose a physics-embedded supervised learning framework for site-specific wireless channel modeling. Rather than treating the channel response as a black-box function, the proposed framework embeds electromagnetic propagation mechanisms into a trainable modal representation. The channel is modeled through source–scatterer–receiver interactions, where scatterer locations and scattering responses (i.e., T-matrices) are learned from measured CSI.
2) We develop an ISAC-assisted initialization and calibration strategy to bridge environmental sensing and communication-domain channel reconstruction. ISAC-induced point clouds provide geometric priors for significant scatterers, and their initial positions are further fine-tuned via backpropagation to compensate for inherent radar sensing errors. CSI samples collected from various receiver locations and beam patterns are employed to optimize the channel representation, ensuring physical consistency.
3) We propose a complexity-tunable multi-scatterer learning mechanism to efficiently capture medium-scale scattering effects. Our model can flexibly span from a first-order Born-type approximation (i.e., single-bounce reflection) to a comprehensive multi-scatterer coupling regime. Specifically, this formulation integrates truncated spherical-wave expansions, multipole-based point-cloud equivalent reshaping, and T-matrix-based scatterer responses [39]. Additionally, a cluster-based extension is developed to enhance scalability in environments with numerous scatterers.

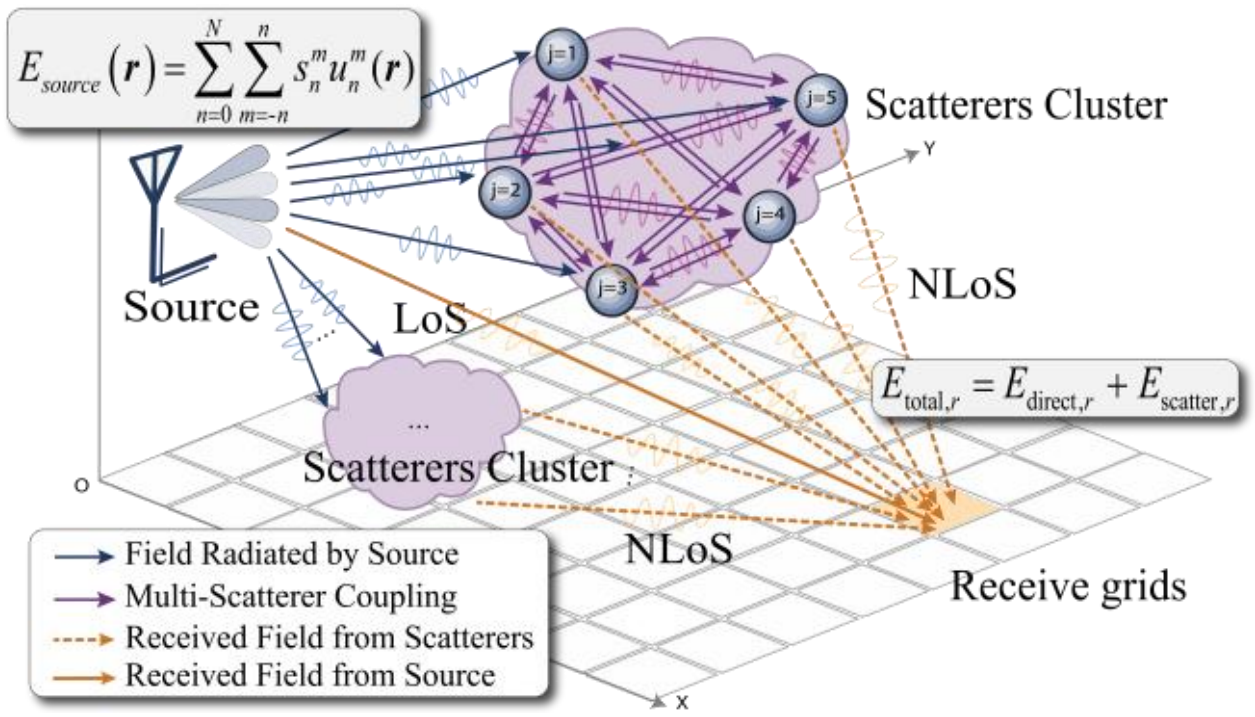

**Fig. 1.** Overview of the multi-scatterer propagation model.

4) Extensive evaluations are conducted to validate the reconstruction and generalization capability of the proposed channel model. Our method is assessed under unseen receiver locations and unseen beam configurations. Simulation results demonstrate the superiority of our method.

Note: Since actual communication measurements typically focus on the overall phase and signal strength rather than fine polarization effects, scalar spherical waves are adopted instead of vector spherical waves to simplify the modeling without loss of main propagation characteristics.

## II. Multi-Scatterer Forward Model

Our model aims to generate continuous radiomaps across the spatial (hereafter referred to as the receiving grid), frequency, and beam domains. Given a specific transmit beam and frequency, the radiomap is obtained by evaluating the complex received field at arbitrary locations. Instead of purely empirical models, we parameterize the local propagation environment utilizing a finite set of learnable discrete scatterers. The physical positions and electromagnetic (EM) scattering responses of these scatterers govern how the transmitted wave is redirected. Once calibrated using sparse measurements, these learnable parameters drive our forward model to predict continuous spatial radiomaps. Fig. 1 illustrates the considered propagation scenario. Formally, the received field at any grid point consists of two parts: the direct incident field representing the line-of-sight (LOS) component and the scattered fields representing the non-line-of-sight (NLOS) multipath components. To ensure EM fidelity, our model incorporates both direct transmitter-to-scatterer interactions and the mutual coupling among multiple scatterers.

### *A. Modal Representation and Translation Operators*

To begin with, we represent radiated and incident fields using a truncated spherical-wave expansion. The standard outgoing spherical-wave basis function and standard regular spherical-wave basis function are respectively denoted as:

$$u_n^m(\boldsymbol{r}) \triangleq h_n^{(1)}(k \cdot r) Y_n^m(\theta,\, \phi) \tag{1a}$$

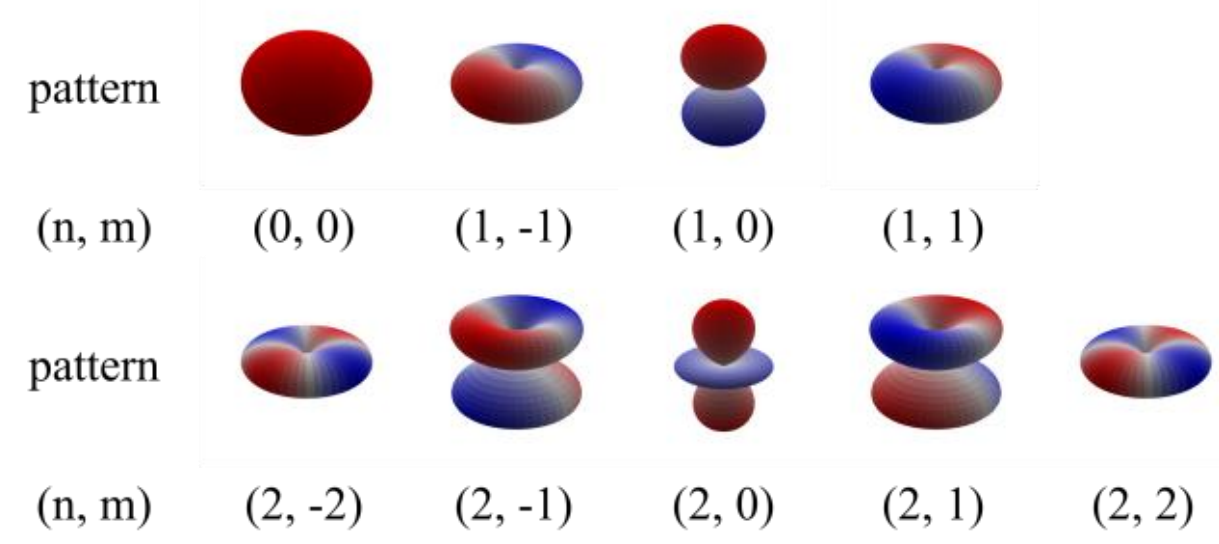

**Fig. 2.** Visualization of low-order spherical-harmonic modes. The color stands for the phase distribution.

$$v_l^p(\boldsymbol{r}) \triangleq j_l(k \cdot r) Y_l^p(\theta,\, \phi) \tag{1b}$$

where $h_n^{(1)}(\cdot)$ and $j_l(\cdot)$ denote the spherical Hankel function of the first kind and spherical Bessel function, respectively, and $Y_n^m(\cdot)$ denotes the spherical harmonic function. The former constitutes the radial component while the latter acts as the angular component. For $\boldsymbol{r} \in \mathbb{R}^3$, let $r = \| \boldsymbol{r} \|$ and $(\theta,\, \phi)$ denote the radial distance and spherical angles of $\boldsymbol{r}$, i.e., elevation angle $\theta \in [0, \pi]$ and azimuth angle $\phi \in [0, 2\pi)$. The wavenumber is $k = 2\pi/\lambda$ with $\lambda$ being the wavelength. For the source-centered expansion (resp. field-centered expansion), define $n \geq 0$ (resp. $l \geq 0$) as the angular quantum number and $m \in \{-n, \dots, n\}$ (resp. $p \in \{-l, \dots, l\}$) as the magnetic quantum number. With these basis functions, the field radiated by the transmitter is given by

$$E_{source}(\boldsymbol{r}) \;=\; \sum_{n=0}^{N} \sum_{m=-n}^{n} s_n^m u_n^m(\boldsymbol{r}), \tag{2}$$

where $\{s_n^m\}$ is the modal coefficient set which encodes the complex radiation pattern of the transmitter.

Fig. 2 visualizes several low-order spherical-harmonic components. Each individual subfigure corresponds to the angular structure associated with one basis $Y_n^m(\theta, \phi)$, while the radial dependence is omitted for clarity. In fact, the radial parts show an asymptotic decaying behavior with respect to the propagation radius, i.e.

$$h_n^{(1)}(k \cdot r) \xrightarrow{r \to \infty} \frac{1}{kr} \exp\left[j\left(k \cdot r - \frac{n\pi}{2} - \frac{\pi}{2}\right)\right] \tag{3a}$$

$$j_l(k \cdot r) \xrightarrow{r \to \infty} \frac{1}{kr} \cos\left(k \cdot r - \frac{l\pi}{2} - \frac{\pi}{2}\right) \tag{3b}$$

*Spherical-wave addition theorem* provides an exact analytical expansion of a spherical wave function centered at one origin using an infinite series of orthogonal spherical wave functions centered at a displacement [40]. Consider two coordinate origins (an original origin and a new origin) separated by a displacement vector $\boldsymbol{d}$, and write $\boldsymbol{r}' = \boldsymbol{r} - \boldsymbol{d}$, where $\boldsymbol{r}'$ is measured from the new origin. When expanding the field near the new origin, $\|\boldsymbol{r}'\| < \|\boldsymbol{d}\|$ holds obviously, so that $\boldsymbol{r}'$ lies in the convergence region guaranteed by the addition theorem. An outgoing mode centered at the original origin can then be expressed as a weighted sum of regular modes centered at the new origin:

$$u_n^m(\boldsymbol{r}) = u_n^m(\boldsymbol{r}; \boldsymbol{d}) = \sum_{l=0}^{L} \sum_{p=-l}^{l} a_{n,m}^{l,p}(\boldsymbol{d}) v_l^p(\boldsymbol{r}') \tag{4}$$

where the translation coefficients $a_{n,m}^{l,p}(\boldsymbol{d})$ depend only on $\boldsymbol{d}$ and the wavenumber $k$. A closed-form expression of $a_{n,m}^{l,p}(\boldsymbol{d})$ is given in (5), where $\overline{X}$ denotes the conjugate of $X$, while

$$\text{Beam 1} = s_0^0[1] \times h_0^{(1)}(kr)Y_0^0(\theta,\phi) + s_1^{-1}[1] \times h_1^{(1)}(kr)Y_1^{-1}(\theta,\phi) + s_1^0[1] \times h_1^{(1)}(kr)Y_1^0(\theta,\phi) + \cdots + s_{\mathrm{n}}^{\mathrm{m}}[1] \times h_{\mathrm{n}}^{(1)}(kr)Y_{\mathrm{n}}^{\mathrm{m}}(\theta,\phi)$$

$$\text{Beam 2} = s_0^0[2] \times h_0^{(1)}(kr)Y_0^0(\theta,\phi) + s_1^{-1}[2] \times h_1^{(1)}(kr)Y_1^{-1}(\theta,\phi) + s_1^0[2] \times h_1^{(1)}(kr)Y_1^0(\theta,\phi) + \cdots + s_{\mathrm{n}}^{\mathrm{m}}[2] \times h_{\mathrm{n}}^{(1)}(kr)Y_{\mathrm{n}}^{\mathrm{m}}(\theta,\phi)$$

$$\text{Beam 3} = s_0^0[3] \times h_0^{(1)}(kr)Y_0^0(\theta,\phi) + s_1^{-1}[3] \times h_1^{(1)}(kr)Y_1^{-1}(\theta,\phi) + s_1^0[3] \times h_1^{(1)}(kr)Y_1^0(\theta,\phi) + \cdots + s_{\mathrm{n}}^{\mathrm{m}}[3] \times h_{\mathrm{n}}^{(1)}(kr)Y_{\mathrm{n}}^{\mathrm{m}}(\theta,\phi)$$

$$\vdots$$

$$\text{Beam i} = s_0^0[i] \times h_0^{(1)}(kr)Y_0^0(\theta,\phi) + s_1^{-1}[i] \times h_1^{(1)}(kr)Y_1^{-1}(\theta,\phi) + s_1^0[i] \times h_1^{(1)}(kr)Y_1^0(\theta,\phi) + \cdots + s_{\mathrm{n}}^{\mathrm{m}}[i] \times h_{\mathrm{n}}^{(1)}(kr)Y_{\mathrm{n}}^{\mathrm{m}}(\theta,\phi)$$

**Fig. 3.** Source-side modal weighting and beam synthesis.

$\begin{pmatrix} n & l & q \\ m & -p & p-m \end{pmatrix}$ and $\begin{pmatrix} n & l & q \\ 0 & 0 & 0 \end{pmatrix}$ are the Wigner-3j coefficients, and

$$C_{nlq} = \delta_{\text{even}}(q+l+n) \cdot \delta_{|p-m|\leq q} \cdot 4\pi(-1)^p \cdot i^{\,q+l-n},$$

where $\delta$ is the indicator function, i.e., $\delta_{\text{even}}(q+l+n) = 1$ (resp. $\delta_{\text{even}}(q+l+n) = 0$) only if $q+l+n$ is even (resp. otherwise), and $\delta_{|p-m|\leq q} = 1$ (resp. $\delta_{|p-m|\leq q} = 0$) only if $|p-m| \leq q$ (resp. otherwise).

By substituting (4) into (2), the coordinate translation from the source at $\boldsymbol{r}_s$ to an arbitrary new expansion center $\boldsymbol{d}$ can be accomplished:

$$E_{inc}(\boldsymbol{r};\boldsymbol{d}) = \sum_{n,m} s_n^m \sum_{l,p} a_{n,m}^{l,p}(\boldsymbol{d}-\boldsymbol{r}_s) v_l^p(\boldsymbol{r}-\boldsymbol{d}) \,, \quad (6)$$

We instantiate the displacement $\boldsymbol{d}$ as the $j$-th scatterer's position $\boldsymbol{r}_j$, and include both the source-induced excitation and the re-radiated fields from other scatterers. The total incident field at the $j$-th scatterer, represented in the regular basis centered at $\boldsymbol{r}_j$, is given as

$$E_{inc}(\boldsymbol{r};\boldsymbol{r}_j) = \sum_{l=0}^{L} \sum_{p=-l}^{l} a_{j,l,p}\, v_l^p(\boldsymbol{r}-\boldsymbol{r}_j), \quad (7)$$

where $\boldsymbol{a}_j \in \mathbb{C}^{(L+1)^2}$ stacks incident modal coefficients $\{a_{j,l,p}\}$ (see (13) for the details). This indicates that the incident wave of the $j$-th scatterer has $(L+1)^2$ modes, with $a_{j,l,p}$ being the incident modal coefficient of the $(l, p)$-th mode.

The modal basis in (1) provides a convenient representation of source-side beam synthesis according to (2). Different transmission beams can be expanded using a set of outgoing spherical-wave modes with the same truncated order, while the corresponding modal coefficients are beam-dependent, as shown in Fig. 3. Specifically, for beam $i$, the source field is characterized by the following coefficient vector

$$\mathbf{s}[i] = [s_0^0[i], s_1^{-1}[i], \cdots, s_n^m[i], \cdots, s_N^N[i]]^T, \quad (8)$$

whose entries specify the weights assigned to different modes. Therefore, beamforming at the source can be regarded as a modal-weight adjusting process: the basis functions are fixed, whereas different coefficient vectors synthesize different beams.

Recall (6), the translation from the source modal coefficients $\boldsymbol{s}$ to the $j$-th scatterer source-induced incident coefficients $\boldsymbol{a}_{sj}$ is linear and can be expressed as

$$\boldsymbol{a}_{sj}[i] = \boldsymbol{H}_{sj}\boldsymbol{s}[i], \quad (9a)$$

where

$$\boldsymbol{H}_{sj} = \begin{bmatrix} a_{0,0}^{0,0}(\boldsymbol{d}_{sj}) & a_{1,-1}^{0,0}(\boldsymbol{d}_{sj}) & \cdots & a_{n,m}^{0,0}(\boldsymbol{d}_{sj}) & \cdots & a_{N,N}^{0,0}(\boldsymbol{d}_{sj}) \\ a_{0,0}^{1,-1}(\boldsymbol{d}_{sj}) & a_{1,-1}^{1,-1}(\boldsymbol{d}_{sj}) & & a_{n,m}^{1,-1}(\boldsymbol{d}_{sj}) & \cdots & a_{N,N}^{1,-1}(\boldsymbol{d}_{sj}) \\ \vdots & & \ddots & \vdots & & \\ a_{0,0}^{l,p}(\boldsymbol{d}_{sj}) & a_{1,-1}^{l,p}(\boldsymbol{d}_{sj}) & & a_{n,m}^{l,p}(\boldsymbol{d}_{sj}) & \cdots & a_{N,N}^{l,p}(\boldsymbol{d}_{sj}) \\ \vdots & \vdots & \cdots & \vdots & \ddots & \vdots \\ a_{0,0}^{L,L}(\boldsymbol{d}_{sj}) & a_{1,-1}^{L,L}(\boldsymbol{d}_{sj}) & & a_{n,m}^{L,L}(\boldsymbol{d}_{sj}) & \cdots & a_{N,N}^{L,L}(\boldsymbol{d}_{sj}) \end{bmatrix} \in \mathbb{C}^{(L+1)^2 \times (N+1)^2} \quad (9b)$$

is the conversion matrix that converts the field equations into coefficients matrix. Each element of $[\boldsymbol{H}_{sj}]_{(\xi),(\zeta)} = a_{n,m}^{l,p}(\boldsymbol{d}_{sj})$, where $\boldsymbol{d}_{sj} = \boldsymbol{r}_j - \boldsymbol{r}_s$, $\boldsymbol{r}_s$ is the source center. Correspondence between matrix column index $\xi$ (resp. row index $\zeta$) and $(n,m)$ (resp. $(l,p)$) is characterized by $\xi = (n+1)^2 + m - n$ (resp. $\zeta = (l+1)^2 + p - l$ ), which reflects the sorting rule for different modes.

*B. Single-Scatterer Scattering*

Consider a single scatterer located at $\boldsymbol{r}_j$. The incident field can be expanded with regular spherical-wave basis centered at $\boldsymbol{r}_j$ and the expansion is truncated at order $L$, which gives the incident coefficient vector $\boldsymbol{a}_j \in \mathbb{C}^{(L+1)^2}$. We model the $j$-th scatterer's field response in the modal domain by a diagonal scattering matrix $\boldsymbol{T}_j \in \mathbb{C}^{(L+1)^2 \times (L+1)^2}$, which mathematically converts incident modal coefficients to outgoing modal coefficients:

$$\boldsymbol{b}_j = \boldsymbol{T}_j \boldsymbol{a}_j, \quad (10)$$

where $\boldsymbol{b}_j \in \mathbb{C}^{(L+1)^2}$ collects the coefficients of the outgoing spherical-wave expansion centered at $\boldsymbol{r}_j$ . $\boldsymbol{T}_j$ reflects the intrinsic scattering properties of the $j$-th scatterer, which is determined by its shape, size, and material distribution. In our learning problem formulated in section III, $\boldsymbol{T}_j$ are treated as unknowns and are estimated from sparse receiver measurements. Given $\boldsymbol{b}_j$, the scattered field at a point $\boldsymbol{r}$ can be

$$a_{n,m}^{l,p}(\boldsymbol{d}) = \sum_{q=|n-l|}^{n+l} C_{n,l,q}\, \overline{Y_q^{p-m}(\theta_d,\phi_d)}\, h_q^{(1)}(k|\boldsymbol{d}|) \begin{pmatrix} n & l & q \\ m & -p & p-m \end{pmatrix} \begin{pmatrix} n & l & q \\ 0 & 0 & 0 \end{pmatrix} \sqrt{\frac{(2n+1)(2l+1)(2q+1)}{4\pi}} \quad (5)$$

expressed as the outgoing spherical-wave expansion

$$E_{scatter}(\boldsymbol{r};\boldsymbol{r_j}) = \sum_{l=0}^{L}\sum_{p=-l}^{l} b_{j,lp}\, u_l^p(\boldsymbol{r}'), \tag{11}$$

where $b_{j,l,p}$ is the $(l,p)$-th entry of $\boldsymbol{b}_j$, $\boldsymbol{r}' = \boldsymbol{r} - \boldsymbol{r}_j$ and $u_l^p(\boldsymbol{r}') = h_l^{(1)}(kr') \cdot Y_l^p(\theta', \phi')$.

*C. Multi-Scatterer Coupling*

Consider an environment with $J$ scatterers located at $\{\boldsymbol{r}_j\}_{j=1}^{J}$. In addition to the source-to-scatterer direct excitation in (9), each scatterer also receives incident waves re-radiated from all other scatterers. Let $\boldsymbol{d}_{ij} = \boldsymbol{r}_j - \boldsymbol{r}_i$ denote the displacement from scatterer $i$ to scatterer $j$. The outgoing modes originated in $\boldsymbol{r}_i$ can be re-expanded as regular modes with respect to $\boldsymbol{r}_j$. This incurs a modal-domain coupling matrix $\boldsymbol{H}_{ij}$, defined as

$$\boldsymbol{H}_{ij} = \begin{bmatrix} a_{0,0}^{0,0}(\boldsymbol{d}_{ij}) & a_{1,-1}^{0,0}(\boldsymbol{d}_{ij}) & \cdots & a_{n,m}^{0,0}(\boldsymbol{d}_{ij}) & \cdots & a_{L,L}^{0,0}(\boldsymbol{d}_{ij}) \\ a_{0,0}^{1,-1}(\boldsymbol{d}_{ij}) & a_{1,-1}^{1,-1}(\boldsymbol{d}_{ij}) & & a_{n,m}^{1,-1}(\boldsymbol{d}_{ij}) & \cdots & a_{L,L}^{1,-1}(\boldsymbol{d}_{ij}) \\ \vdots & & \ddots & & & \vdots \\ a_{0,0}^{l,p}(\boldsymbol{d}_{ij}) & a_{1,-1}^{l,p}(\boldsymbol{d}_{ij}) & & a_{n,m}^{l,p}(\boldsymbol{d}_{ij}) & \cdots & a_{L,L}^{l,p}(\boldsymbol{d}_{ij}) \\ \vdots & \vdots & \cdots & \vdots & \ddots & \vdots \\ a_{0,0}^{L,L}(\boldsymbol{d}_{ij}) & a_{1,-1}^{L,L}(\boldsymbol{d}_{ij}) & & a_{n,m}^{L,L}(\boldsymbol{d}_{ij}) & \cdots & a_{L,L}^{L,L}(\boldsymbol{d}_{ij}) \end{bmatrix} \in \mathbb{C}^{(L+1)^2 \times (L+1)^2} \tag{12}$$

whose entries are determined by the translation coefficients evaluated at $\boldsymbol{d}_{ij}$. Note that $\boldsymbol{H}_{ij}$ is similar to the conversion matrix $\boldsymbol{H}_{sj}$. Since the outgoing and regular expansions are truncated at order $L$, the indices satisfy $0 \leq n, l \leq L$, $m \in \{-n, \dots, n\}$, and $p \in \{-l, \dots, l\}$. Besides, the subscript $(n, m)$ (resp. $(l, p)$) hints on the outgoing mode (resp. regular mode) originated in $\boldsymbol{r}_i$ (resp. $\boldsymbol{r}_j$).

Given the coupling matrix $\boldsymbol{H}_{ij}$, the contribution of scatterer $i$ to the incident coefficients at scatterer $j$ can be written as $\boldsymbol{H}_{ij}\boldsymbol{b}_i$. Combining this with the source direct excitation $\boldsymbol{H}_{sj}\boldsymbol{s}$ in (4) yields the incident coefficient vector $\boldsymbol{a_j}$ at scatterer $j$

$$\boldsymbol{a_j} = \boldsymbol{H}_{sj}\boldsymbol{s} + \sum_{i=1, i\neq j}^{J} \boldsymbol{H}_{ij}\boldsymbol{b}_i, \; j = 1, \dots, J. \tag{13}$$

Using the single-scatterer scattering relation $\boldsymbol{b}_j = \boldsymbol{T}_j\boldsymbol{a}_j$ in (10), a compact form of the multiple-scatterer relation is immediately given by

$$\boldsymbol{b}_j = \boldsymbol{T}_j\left(\boldsymbol{H}_{sj}\boldsymbol{s} + \sum_{i=1, i\neq j}^{J} \boldsymbol{H}_{ij}\, \boldsymbol{b}_i\right), j = 1, \dots, J. \tag{14a}$$

Equation (14b) below is provided as a detailed specification of (14a).

To compactly express the interactions of multiple $\boldsymbol{b}_j$ as calculated in (14a), the scattering coefficient vectors are stacked into $\boldsymbol{b} = [\boldsymbol{b}_1; \dots; \boldsymbol{b}_J] \in \mathbb{C}^{J(L+1)^2}$, and the block-diagonal scattering matrix is defined as $\boldsymbol{T} = blkdiag(\boldsymbol{T}_1, \dots, \boldsymbol{T}_J) \in \mathbb{C}^{J(L+1)^2 \times J(L+1)^2}$. We also construct the block coupling matrix $\boldsymbol{H} \in \mathbb{C}^{J(L+1)^2 \times J(L+1)^2}$ whose $(j, i)$-th block is

$$\boldsymbol{H}[j, i] = \begin{cases} \boldsymbol{H}_{ij}, & i \neq j \\ 0, & i = j \end{cases}, \tag{15}$$

and $\boldsymbol{H}_s \triangleq [\boldsymbol{H}_{s1}; \dots; \boldsymbol{H}_{sJ}] \in \mathbb{C}^{J(L+1)^2 \times (N+1)^2}$ stacks the source-to-scatterer conversion matrices. Then (14) can be written as the global linear system:

$$(\boldsymbol{I} - \boldsymbol{T}\boldsymbol{H})\boldsymbol{b} = \boldsymbol{T}\boldsymbol{H}_s\boldsymbol{s}. \tag{16a}$$

whose detailed specification is given by (16b). Given the transmitter coefficients vector $\boldsymbol{s}$ and the scattering matrices $\{\boldsymbol{T}_j\}$, $\boldsymbol{b}$ can be solved from (14) (as discussed in Section E), and the received field can then be computed using the direct-field and scattered-field, which will be explained in the next subsection.

---

$$\begin{bmatrix} b_{j,0,0} \\ b_{j,1,-1} \\ \vdots \\ b_{j,l,p} \\ \vdots \\ b_{j,L,L} \end{bmatrix} = \underbrace{\begin{bmatrix} t_{j,0,0} & 0 & \cdots & 0 \\ 0 & t_{j,1,-1} & & 0 \\ \vdots & & \ddots & \vdots \\ 0 & 0 & \cdots & t_{j,L,L} \end{bmatrix}}_{\boldsymbol{T}_j \in \mathbb{C}^{(L+1)^2 \times (L+1)^2}} \left( \underbrace{\begin{bmatrix} a_{0,0}^{0,0}(\boldsymbol{d}_{sj}) & a_{1,-1}^{0,0}(\boldsymbol{d}_{sj}) & \cdots & a_{n,m}^{0,0}(\boldsymbol{d}_{sj}) & \cdots & a_{N,N}^{0,0}(\boldsymbol{d}_{sj}) \\ a_{0,0}^{1,-1}(\boldsymbol{d}_{sj}) & a_{1,-1}^{1,-1}(\boldsymbol{d}_{sj}) & & a_{n,m}^{1,-1}(\boldsymbol{d}_{sj}) & \cdots & a_{N,N}^{1,-1}(\boldsymbol{d}_{sj}) \\ \vdots & & \ddots & & & \vdots \\ a_{0,0}^{l,p}(\boldsymbol{d}_{sj}) & a_{1,-1}^{l,p}(\boldsymbol{d}_{sj}) & & a_{n,m}^{l,p}(\boldsymbol{d}_{sj}) & \cdots & a_{N,N}^{l,p}(\boldsymbol{d}_{sj}) \\ \vdots & \vdots & \cdots & \vdots & \ddots & \vdots \\ a_{0,0}^{L,L}(\boldsymbol{d}_{sj}) & a_{1,-1}^{L,L}(\boldsymbol{d}_{sj}) & & a_{n,m}^{L,L}(\boldsymbol{d}_{sj}) & \cdots & a_{N,N}^{L,L}(\boldsymbol{d}_{sj}) \end{bmatrix}}_{\boldsymbol{H}_{sj} \in \mathbb{C}^{(L+1)^2 \times (N+1)^2}} \begin{bmatrix} s_0^0 \\ s_1^{-1} \\ \vdots \\ s_n^m \\ \vdots \\ s_N^N \end{bmatrix} \right.$$

$$\left. + \sum_{\boldsymbol{i=1, i\neq j}}^{\boldsymbol{J}} \underbrace{\begin{bmatrix} a_{0,0}^{0,0}(\boldsymbol{d}_{ij}) & a_{1,-1}^{0,0}(\boldsymbol{d}_{ij}) & \cdots & a_{n,m}^{0,0}(\boldsymbol{d}_{ij}) & \cdots & a_{L,L}^{0,0}(\boldsymbol{d}_{ij}) \\ a_{0,0}^{1,-1}(\boldsymbol{d}_{ij}) & a_{1,-1}^{1,-1}(\boldsymbol{d}_{ij}) & & a_{n,m}^{1,-1}(\boldsymbol{d}_{ij}) & \cdots & a_{L,L}^{1,-1}(\boldsymbol{d}_{ij}) \\ \vdots & & \ddots & & & \vdots \\ a_{0,0}^{l,p}(\boldsymbol{d}_{ij}) & a_{1,-1}^{l,p}(\boldsymbol{d}_{ij}) & & a_{n,m}^{l,p}(\boldsymbol{d}_{ij}) & \cdots & a_{L,L}^{l,p}(\boldsymbol{d}_{ij}) \\ \vdots & \vdots & \cdots & \vdots & \ddots & \vdots \\ a_{0,0}^{L,L}(\boldsymbol{d}_{ij}) & a_{1,-1}^{L,L}(\boldsymbol{d}_{ij}) & & a_{n,m}^{L,L}(\boldsymbol{d}_{ij}) & \cdots & a_{L,L}^{L,L}(\boldsymbol{d}_{ij}) \end{bmatrix}}_{\boldsymbol{H}_{ij} \in \mathbb{C}^{(L+1)^2 \times (L+1)^2}} \begin{bmatrix} b_{i,0,0} \\ b_{i,1,-1} \\ \vdots \\ b_{i,l,p} \\ \vdots \\ b_{i,L,L} \end{bmatrix} \right) \tag{14b}$$

$$\left( \begin{bmatrix} \boldsymbol{I} & \boldsymbol{0} & \dots & \boldsymbol{0} \\ \boldsymbol{0} & \boldsymbol{I} & \dots & \boldsymbol{0} \\ \vdots & \vdots & \ddots & \vdots \\ \boldsymbol{0} & \boldsymbol{0} & \dots & \boldsymbol{I} \end{bmatrix} - \begin{bmatrix} \boldsymbol{0} & \boldsymbol{T_1 H_{21}} & \dots & \boldsymbol{T_1 H_{J1}} \\ \boldsymbol{T_2 H_{12}} & \boldsymbol{0} & \dots & \boldsymbol{T_2 H_{J2}} \\ \vdots & \vdots & \ddots & \vdots \\ \boldsymbol{T_J H_{1J}} & \boldsymbol{T_J H_{2J}} & \dots & \boldsymbol{0} \end{bmatrix} \right) \begin{bmatrix} \boldsymbol{b_1} \\ \boldsymbol{b_2} \\ \vdots \\ \boldsymbol{b_J} \end{bmatrix} = \begin{bmatrix} \boldsymbol{T_1 H_{s1} s} \\ \boldsymbol{T_2 H_{s2} s} \\ \vdots \\ \boldsymbol{T_J H_{sJ} s} \end{bmatrix} \tag{16b}$$

*D. Receiver-Side Field*

Consider a receiver located at $\boldsymbol{r}_r$. The received field consists of a line-of-sight path from the transmitter and a scattered component from all scatterers:

$$E_{total,r} = E_{direct,r} + E_{scatter,r} \tag{17}$$

For the direct component, the transmitter field in (2) gives:

$$E_{direct,r} = \sum_{n=0}^{N} \sum_{m=-n}^{n} s_n^m u_n^m(\boldsymbol{r}_{sr}). \tag{18}$$

where $\boldsymbol{r}_{sr} = \boldsymbol{r}_r - \boldsymbol{r}_s$ . For the scattered component, each scatterer $j$ contributes an outgoing spherical-wave expansion centered at $\boldsymbol{r}_j$. Using the scattering coefficients $\{\boldsymbol{b}_j\}_{j=1}^{J}$, we obtain

$$E_{scatter,r} = \sum_{j=1}^{J} \sum_{l=0}^{L} \sum_{p=-l}^{l} b_{j,lp}\, u_l^p(\boldsymbol{r}_{jr}). \tag{19}$$

where $\boldsymbol{r}_{jr} = \boldsymbol{r}_r - \boldsymbol{r}_j$. Once the global scattering coefficient $\boldsymbol{b}$ is obtained, the received field at any receiver location can be computed through (17)-(19).

*E. Iterative Solution for Large Matrix Inversion*

In the learning process introduced in Section III, equation (16) must be solved at each epoch. The most straightforward approach for solving (16) is to directly compute the inverse of the matrix $(\boldsymbol{I} - \boldsymbol{TH})$ . However, this incurs a significant computational burden, especially when the matrix $(\boldsymbol{I} - \boldsymbol{TH})$ is large. To achieve an affordable complexity, we introduce iterative methods to approximately solve (16). The resulting iteration admits a physical interpretation through the Born-series expansion: each additional iteration introduces higher-order inter-scatterer coupling contributions. Therefore, the number of iterations provides a controllable tradeoff between computational complexity and the order of multiple-scattering interactions included in the forward solution.

In (16), the collective scattering coefficient vector $\boldsymbol{b} = [\boldsymbol{b}_1; \dots; \boldsymbol{b}_J]$. Denoting $\boldsymbol{M} \triangleq \boldsymbol{TH}$ as the mutual coupling matrix, which has full-zero diagonal blocks and non-zero off-diagonal blocks, $\boldsymbol{M}$ can be separated into its strictly block-lower and strictly block-upper triangular parts, respectively denoted by $\boldsymbol{L}$ and $\boldsymbol{U}$, so that $\boldsymbol{M} = \boldsymbol{L} + \boldsymbol{U}$. In block form,

$$[\boldsymbol{L} + \boldsymbol{U}]_{j,i} = \begin{cases} \boldsymbol{T}_j \boldsymbol{H}_{ij}, & i \neq j \\ 0, & i = j \end{cases}. \tag{20}$$

where the $(j, i)$-th block represents the contribution of the re-radiated field from scatterer $i$ to the outgoing scattering coefficients of scatterer $j$. With $\boldsymbol{c} \triangleq \boldsymbol{TH}_s\boldsymbol{s}$, (16) can be written as $[\boldsymbol{I} - (\boldsymbol{L} + \boldsymbol{U})]\boldsymbol{b} = \boldsymbol{c}$, which is suitable for iterative inversion due to the separable identity diagonal part. A natural choice is the Jacobi iteration, which is

$$\boldsymbol{b}^{(k+1)} = (\boldsymbol{L} + \boldsymbol{U}) \cdot \boldsymbol{b}^{(k)} + \boldsymbol{c}, \tag{21a}$$

Jacobi iteration is well suited for parallel implementation. Expanding (21a) further gives

$$\begin{aligned} \boldsymbol{b}^{(k+1)} &= (\boldsymbol{L} + \boldsymbol{U}) \cdot \left[(\boldsymbol{L} + \boldsymbol{U}) \cdot \boldsymbol{b}^{(k-1)} + \boldsymbol{c}\right] + \boldsymbol{c} \\ &= (\boldsymbol{L} + \boldsymbol{U})^{k+1} \cdot \boldsymbol{b}^{(0)} + \textstyle\sum_{i=0}^{k} (\boldsymbol{L} + \boldsymbol{U})^i \cdot \boldsymbol{c} \end{aligned} \tag{21b}$$

When $\boldsymbol{b}^{(0)}$ is initialized as a zero vector, the term $(\boldsymbol{L}+\boldsymbol{U})^i\boldsymbol{c}$ is interpreted as contribution associated with the $i$-th inter-scatterer interaction. Consequently, after $k+1$ Jacobi iterations the resulting scattering coefficients is essentially the $(k+1)$-th order Born approximation, provide that $\boldsymbol{c}$ is the first-order scattering contribution. When the spectral radius $\rho(\boldsymbol{L} + \boldsymbol{U}) < 1$, the Jacobian iteration rapidly converges, which mirrors the fact that the energy of EM wave diminishes after multiple scatterings. In per-scatterer block form, the Jacobi update becomes

$$\boldsymbol{b}_j^{(k+1)} = \textstyle\sum_{i \neq j} \boldsymbol{T}_j \boldsymbol{H}_{ij} \boldsymbol{b}_i^{(k)} + \boldsymbol{T}_j \boldsymbol{H}_{sj} \boldsymbol{s}, j = 1..J, \tag{22}$$

Although Jacobi iteration is highly parallelizable, it does not use the most recently updated block solutions within the same iteration, which may lead to slow convergence.

Alternatively, Gauss-Seidel (GS) iteration, which utilizes the most recent result within the current iteration, is known to provide a faster convergence speed. The matrix form of GS iteration is given as

$$(\boldsymbol{I} - \boldsymbol{L}) \cdot \boldsymbol{b}^{(k+1)} = \boldsymbol{U} \cdot \boldsymbol{b}^{(k)} + \boldsymbol{c}, \tag{23}$$

In term of each block, the GS update becomes

$$\begin{aligned} \boldsymbol{b}_j^{(k+1)} &= \textstyle\sum_{i<j} \boldsymbol{T}_j \boldsymbol{H}_{ij} \boldsymbol{b}_i^{(k+1)} + \\ &\textstyle\sum_{i>j} \boldsymbol{T}_j \boldsymbol{H}_{ij} \boldsymbol{b}_i^{(k)} + \boldsymbol{T}_j \boldsymbol{H}_{sj} \boldsymbol{s},\ j = 1..J \end{aligned} \tag{24}$$

Compared to Jacobian method, GS replaces the "earlier" blocks ( $i < j$ ) by their newly updated values, resulting in faster convergence.

Nevertheless, when the scatterers are strongly coupled or when the number of scatterers becomes large, both Jacobian method and GS iterations may diverge due to the non-diagonal-dominance of $\boldsymbol{I} - \boldsymbol{M}$ . Fortunately, the successive over-relaxation (SOR) can be employed to mitigate this problem. The specific form of SOR is read as

$$(\boldsymbol{I} - \omega\boldsymbol{L})\boldsymbol{b}^{(k+1)} = [(1 - \omega)\boldsymbol{I} + \omega\boldsymbol{U}] \cdot \boldsymbol{b}^{(k)} + \omega\boldsymbol{c}. \tag{25}$$

where $\omega \in (0,2)$ is the relaxation factor.

*F. Inter-Cluster Approximation*

In practical scenarios, each large-scale scatterer will be sensed as a cluster of point clouds due to multiple scattering centers in ISAC system. The mutual coupling inside each cluster is more significant than that between different clusters, motivating an inter-cluster approximation. To elaborate, in the first inter-cluster update, both the inter- and intra-cluster coupling are considered, and in subsequent iterations, only the intra-coupling inside each cluster is calculated. This can effectively reduce the computational complexity. For clarity, consider the example where the scatterers are divided into two clusters, denoted by $\mathcal{C}_1$ and $\mathcal{C}_2$ , The global linear system vectors in (16a) are correspondingly partitioned as $\boldsymbol{b} = [\boldsymbol{b}_1\ \boldsymbol{b}_2]^{\mathrm{T}}$ and $\boldsymbol{c} \triangleq \boldsymbol{TH}_s\boldsymbol{s} = [\ \boldsymbol{c}_1\ \ \boldsymbol{c}_2\ ]^{\mathrm{T}}$, where $\boldsymbol{b}_g$ and $\boldsymbol{c}_g$ collect the coefficients associated with cluster $\mathcal{C}_g$, respectively. The global mutual coupling matrix can be partitioned as

$$\boldsymbol{M} = \begin{bmatrix} \boldsymbol{M}_{11} & \boldsymbol{M}_{12} \\ \boldsymbol{M}_{21} & \boldsymbol{M}_{22} \end{bmatrix},$$

where each diagonal matrix (resp. non-diagonal matrix) represents the intra-cluster coupling (resp. inter-cluster coupling). The mutual coupling matrix is decomposed into the inter-cluster part and the intra-cluster part as

$$\boldsymbol{M}_{inter} = \begin{bmatrix} 0 & \boldsymbol{M}_{12} \\ \boldsymbol{M}_{21} & 0 \end{bmatrix}$$

and

$$\boldsymbol{M}_{intra} = \begin{bmatrix} \boldsymbol{M}_{11} & 0 \\ 0 & \boldsymbol{M}_{22} \end{bmatrix},$$

respectively.

As per (21a), $\boldsymbol{b}_1^{(1)}$ and $\boldsymbol{b}_2^{(1)}$ are respectively initialized as $\boldsymbol{c}_1$ and $\boldsymbol{c}_2$, indicating each cluster is excited only by the source field. The scattered wave generated by each cluster after the first source-driven update is then translated to the other cluster. This gives an additional inter-cluster incident excitation

$$\Delta\boldsymbol{c} = \begin{bmatrix} \Delta\boldsymbol{c}_1 \\ \Delta\boldsymbol{c}_2 \end{bmatrix} = \boldsymbol{M}_{inter}\boldsymbol{b}^{(1)} = \begin{bmatrix} \boldsymbol{M}_{12}\boldsymbol{c}_2 \\ \boldsymbol{M}_{21}\boldsymbol{c}_1 \end{bmatrix}, \tag{26}$$

where $\Delta\boldsymbol{c}_1$ (resp. $\Delta\boldsymbol{c}_2$) represents the excitation from cluster $\mathcal{C}_2$ to cluster $\mathcal{C}_1$ (resp. cluster $\mathcal{C}_1$ to cluster $\mathcal{C}_2$). After this inter-cluster excitation, subsequent mutual coupling process is restricted within each cluster. Thus, the intra-cluster system is formulated as

$$(\boldsymbol{I} - \boldsymbol{M}_{intra})\boldsymbol{b} = \boldsymbol{c} + \Delta\boldsymbol{c}, \tag{27a}$$

which can be further expanded as

$$\begin{bmatrix} \boldsymbol{I} - \boldsymbol{M}_{11} & \boldsymbol{0} \\ \boldsymbol{0} & \boldsymbol{I} - \boldsymbol{M}_{22} \end{bmatrix} \begin{bmatrix} \boldsymbol{b}_1 \\ \boldsymbol{b}_2 \end{bmatrix} = \begin{bmatrix} \boldsymbol{c}_1 + \Delta\boldsymbol{c}_1 \\ \boldsymbol{c}_2 + \Delta\boldsymbol{c}_2 \end{bmatrix}, \tag{27b}$$

and thus, can be conveniently solved by the SOR method.

## III. Inverse Problem and Cross-Domain Generalization Applications

### *A. Formulation of Multi-Scatterer Parameter Inversion*

To predict the radiomap distributed on denser/new spatial grids, or under the illumination of new beam patterns, the property of the environment, i.e., scattering matrix $\boldsymbol{T}$ and scatterers' positions $\boldsymbol{R}$ should be fully known. Once these parameters are available, the forward model (16)-(19) can be evaluated on arbitrary receiver grids to generate the corresponding radiomap. In practice, however, only sparse CSI measurements collected over a limited number of receiver locations and beam patterns are available. Therefore, we formulate an inverse problem to learn the equivalent scatterer parameters from sparse field observations, which is also known as parameter inversion.

In our model, the scattering matrices and positions of the scatterers determine the propagation behavior of a specific scene. To elaborate, the position of the $j$-th virtual scatterer is parameterized by $\boldsymbol{r}_j = \bar{\boldsymbol{r}}_j + \Delta\boldsymbol{r}_j$, where $\bar{\boldsymbol{r}}_j$ denotes the original scatterers' location (anchor) and $\Delta\boldsymbol{r}_j$ is a learnable offset. By imposing this subtle offset $\Delta\boldsymbol{r}_j$, the virtual scatterers are forced to lie around the anchors. The reason of the introduced offsets is that, the ISAC point clouds represent the scattering center of the environmental objects; however, these anchor locations may not be perfectly consistent with their geometric center. In addition, the ISAC cloud points always reflect a monostatic RCS of a scatterer instead of multi-static RCS. Therefore, the angular dependent responses are attributed to the scattering matrix $\boldsymbol{T}_j$, which is another learnable parameter set.

Assume the complex field is measured at a small set of receiver locations, and measurements are stacked into a vector $\boldsymbol{y}_{meas} \in \mathbb{C}^Q$, where $Q$ is the number of measurement points. In the context of wireless communication, the measurements refer to the CSI measured at multiple geographical receiving grids. In addition, the phaseless data [41], such as Reference Signal Receiving Power (RSRP) or CSI magnitude, are easier to obtain, despite that this potentially increases the ill-posedness of the inversion.

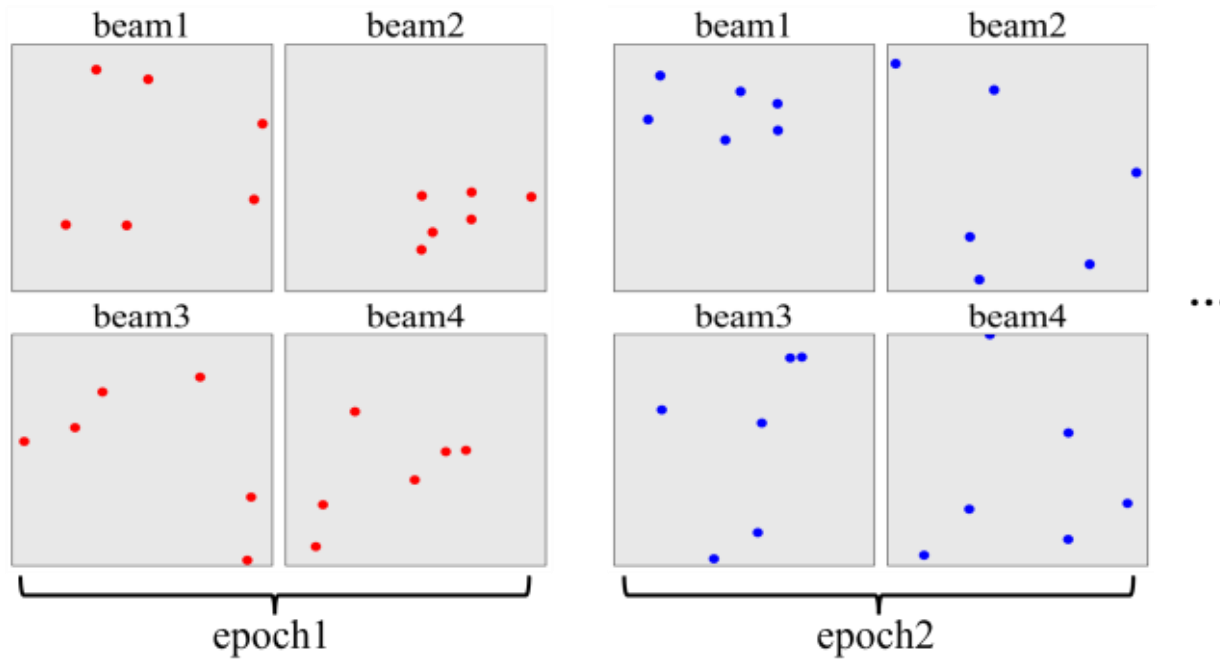

**Fig. 4.** Schematic illustration of epoch-wise random sampling on the sparse receiver plane. Different colors denote the receiver points selected at different epochs.

For a given set of scattering matrices $\{\boldsymbol{T}_j\}$ and anchor $\{\boldsymbol{r}_j\}$ under illumination of a beam determined by $\boldsymbol{s}$, we first solve the coupled multi-scatterer system (16), and then evaluate the receiver-side field using (17)-(19). This procedure is formularized as

$$\boldsymbol{y} = \mathcal{H}[\boldsymbol{b}] = \mathcal{H}(\mathcal{F}[\boldsymbol{T}, \boldsymbol{R}; \boldsymbol{s}]) \in \mathbb{C}^Q, \tag{28}$$

where $\mathcal{F}[\cdot]$ is the backward mapping that outputs $\boldsymbol{b}$ by solving (16), and $\mathcal{H}[\cdot]$ is the forward operator corresponding to (17)-(19).

Given sparse measurement $\boldsymbol{y}_{meas}$ and fixed anchor positions $\bar{\boldsymbol{R}} = \{\bar{r}_j\}$, two variables, namely $\boldsymbol{T}$ and $\Delta\boldsymbol{R} = \{\Delta\boldsymbol{r}_j\}$, are optimized by solving the following learning problem:

$$(\boldsymbol{T}^\star, \Delta\boldsymbol{R}^\star) \in \underset{\boldsymbol{T},\Delta\boldsymbol{R}}{argmin} \frac{\sum_{q=1}^{Q}\left|y_q(\boldsymbol{T},\Delta\boldsymbol{R};\bar{\boldsymbol{R}}) - y_{q,meas}\right|^2}{\sum_{q=1}^{Q}\left|y_{q,meas}\right|^2}. \tag{29a}$$

$$s.t.\ \boldsymbol{T}_j{}^{\boldsymbol{H}}\boldsymbol{T}_j \preceq \gamma\boldsymbol{I}, \forall j. \tag{29b}$$

$$\left\|\Delta\boldsymbol{r}_j\right\|_2 < \rho, \forall j \tag{29c}$$

where $y_q(\boldsymbol{T}, \Delta\boldsymbol{R}; \bar{\boldsymbol{R}})$ denotes the predicted CSI of the $q$-th *training* sample in the "grid × beam" space, and $y_{q,meas}$ represents the $q$-th CSI measurement. Note that $y_q$ is a function of $\boldsymbol{T}$ and $\Delta\boldsymbol{R}$, conditioned on $\bar{\boldsymbol{R}}$**.** A prescribed bound $\gamma$ is used to control the magnitude of $\boldsymbol{T}_j$. The constraint in (30b) limits the energy of each scattering matrix and prevents the scatterers from overfitting with unrealistically strong responses. The position offsets are also restricted within a radius $\rho$ around anchors. In the simulation, we use scattering matrix $\boldsymbol{T}_{tar}$ and reference scatterers positions $\boldsymbol{R}_{tar}$ in the target model to derive reference measurement $\boldsymbol{y}_{tar}$, i.e., $\boldsymbol{y}_{tar} = \mathcal{H}(\mathcal{F}[\boldsymbol{T}_{tar}, \boldsymbol{R}_{tar}; \boldsymbol{s}])$, and $y_{q,meas}$ is a copy of $y_{q,tar}$ but imposed with additive Gaussian noise to simulate real scenarios. The model is trained using mini-batches of receiving points. For each beam, a random subset of the receiver grid is selected in each training epoch to form a mini-batch, as illustrated in Fig. 4.

In real-world channels, impairments such as Doppler shifts and thermal noise accumulate as rapid phase drift, making CSI phase unstable for long-term use. The magnitude information, on the other hand, is dominated by large-scale fading and is relatively stable. Hence, a robust long-term channel metric is the CSI magnitude, and an alternative loss function discarding phase information can be formulated as

$$(\boldsymbol{T}^{\star}, \Delta\boldsymbol{R}^{\star}) \in \underset{\boldsymbol{T},\Delta\boldsymbol{R}}{argmin} \frac{\sum_{q=1}^{Q} \left| |y_q(\boldsymbol{T},\Delta\boldsymbol{R};\bar{\boldsymbol{R}})| - |y_{q,meas}| \right|^2}{\sum_{q=1}^{Q} |y_{q,meas}|^2}. \tag{29d}$$

*B. Order Reduction via Multipole Virtualization*

Based on the spherical-wave addition theorem (4) [40], even a single-mode multipole translated from its original center can excite a superposition of multiple modes. In addition, high-order spherical-wave harmonics are poorly radiative [42]. These spark the idea of using a large number of low-order-truncated multipoles with certain displacements to represent a small number of high-order-truncated multipoles. To elaborate, assuming that the original scene contains $J_1$ real scatterers truncated at order $L_1$. In fact, these scatterers can be approximated by $J_2$ virtual scatterers located at $\left\{ \boldsymbol{r}_j^{(low)} = \bar{\boldsymbol{r}}_j^{(low)} + \Delta\boldsymbol{r}_j^{(low)} \right\}_{j=1}^{J_2}$, where $\bar{\boldsymbol{r}}_j^{(low)}$ denotes the anchor position replicated from the original real scatterer locations, $\Delta\boldsymbol{r}_j^{(low)}$ is a small and learnable offset. In this case, each anchor location is surrounded by several virtual scatterers.

For the quantitative details, the truncated order of each scatterer is $L_2$, where $J_2 > J_1$, $L_2 < L_1$, and $J_2(L_2+1)^2 << J_1(L_1+1)^2$. We optimize the virtual scattering matrix $\boldsymbol{T}^{(low)} = blkdiag\left(\boldsymbol{T}_1^{(low)}, \ldots, \boldsymbol{T}_{J_2}^{(low)}\right) \in \mathbb{C}^{J_2(L_2+1)^2 \times J_2(L_2+1)^2}$ and virtual scatterers offset $\Delta\boldsymbol{R}^{(low)} = \{\Delta\boldsymbol{r}_j^{(low)}\}_{j=1}^{J_2}$. The forward model has a similar form as (16) but with different virtual geometry and truncated order:

$$(\boldsymbol{I} - \boldsymbol{T}^{(low)}\boldsymbol{H}^{(low)})\boldsymbol{b}^{(low)} = \boldsymbol{T}^{(low)}\boldsymbol{H}_S^{(low)}\boldsymbol{s}\,. \tag{30}$$

The predicted radiomap samples are obtained via

$$\boldsymbol{y}^{(low)} = \mathcal{H}\left[\boldsymbol{b}^{(low)}\right] = \mathcal{H}\left(\mathcal{F}\left[\boldsymbol{T}^{(low)}, \boldsymbol{R}^{(low)}\right]\right) \in \mathbb{C}^Q \tag{31}$$

At last, the learning problem is accordingly formulated as

$$(\boldsymbol{T}^{(low)\star}, \Delta\boldsymbol{R}^{(low)\star}) \in$$

$$\underset{\boldsymbol{T}^{(low)},\Delta\boldsymbol{R}^{(low)}}{argmin} \frac{\sum_{q=1}^{Q} \left| y_q^{(low)}(\boldsymbol{T}^{(low)},\Delta\boldsymbol{R}^{(low)};\bar{\boldsymbol{R}}) - y_{q,meas}^{(low)} \right|^2}{\sum_{q=1}^{Q} \left| y_{q,meas}^{(low)} \right|^2}, \tag{32a}$$

$$s.t.\ (\boldsymbol{T}_j^{(low)})^{\boldsymbol{H}}\boldsymbol{T}_j^{(low)} \preceq \gamma\boldsymbol{I}, \forall j, \tag{32b}$$

$$\left\| \Delta\boldsymbol{r}_j^{(low)} \right\|_2 < \rho, \forall j\,. \tag{32c}$$

This formulation shifts model complexity from "higher-order modes per scatterer" to "more scatterers with lower-order modes", thus reducing problem complexity and ill-posedness.

*C. Beam-wise Radiomap Prediction*

To adapt the trained model to unseen beam patterns, we develop a beam expansion method that synthesizes an arbitrary beam in the modal subspace. In particular, notice that an arbitrary beam pattern can be written as the superposition of multiple low-order spherical harmonics, that is,

$$\begin{aligned} F_i(\theta,\varphi;r_0) &= \textstyle\sum_n \sum_m s[i]_n^m h_n^{(1)}(kr_0) Y_n^m(\theta,\varphi) \\ &= \textstyle\sum_n \sum_m a[i]_n^m(r_0) Y_n^m(\theta,\varphi) \end{aligned} \tag{33}$$

where $a[i]_n^m(r_0) = s[i]_n^m h_n^{(1)}(kr_0)$ denotes the spherical-harmonic coefficient of the $i$-th beam evaluated on a fixed far-field sphere with radius $r_0$. $a[i]_n^m(r_0)$ can be determined by

$$a[i]_n^m(r_0) = \int_0^{2\pi} \int_0^{\pi} F_i(\theta,\varphi;r_0) Y_n^{m*}(\theta,\varphi) sin\theta d\theta\, d\varphi \tag{34}$$

By using the orthogonality of spherical harmonics, the incident coefficients are calculated by

$$s[i]_n^m = a[i]_n^m(r_0)/h_n^{(1)}(kr_0) \tag{35}$$

which implies that the transmitter coefficients of the $i$-th beam can be easily obtainable.

Upon (33)-(35), even if a beam pattern is unseen in the training set, it can be synthesized by projecting its angular pattern onto the spherical-harmonic basis to obtain the corresponding source modal coefficients, thereby enabling beam-wise generalization to unseen beam patterns. For the $\kappa$ training beams, the corresponding coupled systems can be stacked and solved jointly as follows:

$$(\boldsymbol{I} - \boldsymbol{TH})\boldsymbol{B} = \boldsymbol{TH}_s\boldsymbol{S} \tag{36}$$

where

$$\boldsymbol{B} = [\boldsymbol{b}_1, \boldsymbol{b}_2, \ldots \boldsymbol{b}_\kappa] \in \mathbb{C}^{J(L+1)^2 \times \kappa} \tag{37}$$

and

$$\boldsymbol{S} = [\boldsymbol{s}_1, \boldsymbol{s}_2, \ldots \boldsymbol{s}_\kappa] \in \mathbb{C}^{(N+1)^2 \times \kappa} \tag{38}$$

At last, the beam generalization matrix can be numerically expressed by

$$\boldsymbol{Q} \triangleq \boldsymbol{B}\boldsymbol{S}^{-1} = (\boldsymbol{I} - \boldsymbol{TH})^{-1}\boldsymbol{TH}_s \tag{39}$$

which can be further used for predicting the scattering coefficients under a new transmitter coefficient, that is

$$\boldsymbol{b}_{new} = \boldsymbol{Q}\boldsymbol{s}_{new} \tag{40}$$

Also note that $\boldsymbol{S}^{-1}$ should be replaced by the pseudo-inverse form of the beam generalization matrix as long as $\boldsymbol{S}$ is not a square matrix.

*D. Channel Frequency Response (CFR) Interpolation*

Channel estimation is fundamental in wireless communication systems. In practice, reference signals (RS) are allocated to a sparse subset of resource elements (REs) to directly acquire the channel responses at these specific locations, while the responses at the remaining REs are subsequently obtained via interpolation. Under this logic, our proposed model is trained on a limited set of discrete frequencies and then evaluated at unobserved intermediate frequencies. Specifically, we extend the data dimension from "grid × beam" space to "grid × beam × frequency" space, therefore (32) is modified as

$$(\boldsymbol{T}(f_\upsilon)^{(low)\star}, \Delta\boldsymbol{R}^{(low)\star}) \in$$

$$\underset{\boldsymbol{T}(f_\upsilon)^{(low)},\Delta\boldsymbol{R}^{(low)}}{argmin} \frac{\sum_{\upsilon=1}^{F}\sum_{q=1}^{Q} \left| y_q^{(low)}(\boldsymbol{T}(f_\upsilon)^{(low)},\Delta\boldsymbol{R}^{(low)};\bar{\boldsymbol{R}}) - y_{f_\upsilon,q,meas}^{(low)} \right|^2}{\sum_{\upsilon=1}^{F}\sum_{q=1}^{Q} \left| y_{f_\upsilon,q,meas}^{(low)} \right|^2}, \tag{41}$$

where $y_q^{(low)}(\boldsymbol{T}(f_\upsilon)^{(low)}, \Delta\boldsymbol{R}^{(low)}; \bar{\boldsymbol{R}})$ denotes the predicted CSI generated by the low-order forward model at the $q$-th "grid × beam" sample and the $\nu$-th frequency, $y_{f_\upsilon,q,meas}^{(low)}$ is the measured CSI sample, $f_\upsilon$ is the $\upsilon$-th frequency point, and $F$ is the number of frequency points involved in the training process. In (41), we assume that the scatterers share the same position with dispersive (frequency-dependent) scattering T-matrices. After the training process, the T-matrices for any intermediate frequency point can be interpolated (e.g. spline) over those of training frequencies. Finally, the CFR at any intermediate frequency point can be obtained.

## IV. Simulation Results

In this section, we provide evaluation results of our proposed method. The experiments are designed to validate four aspects of the proposed method: the convergence behavior of the iterative forward solver, the reconstruction accuracy of the lower-order equivalent model, the effectiveness of the inter-cluster approximation, and the generalization capability over unseen receiver locations, beam patterns, and frequencies.

### *A. Simulation Setup*

Considering a narrowband wireless propagation scenario, in which all geometric dimensions are normalized by the wavelength. The carrier frequency $f$ can be chosen arbitrarily without changing the normalized simulation results. The reference scene size is set to $A = 100\lambda$. A source is placed at $\boldsymbol{S} = [0, 0, 0.9A]$, radiating $\kappa = 8$ beams represented by truncated outgoing spherical-wave coefficients. For the ground-truth environment, $N_s = 20$ scatterers are generated in a cubic region inside the scene, see Fig. 5. We construct two scenarios: one with all scatterers grouped into a single cluster, and the other with scatterers partitioned into two distinct clusters. The two-cluster scene is used to verify the proposed inter-cluster approximation. The truncation order of the ground-truth scatterers is set to $L_1 = 3$, whereas that of its lower-order equivalent model is fixed as $L_2 = 1$ which corresponds to dipole-level virtual scatterers. Each original scatterer is replaced by two lower-order virtual scatterers placed near its original location in the equivalent model, resulting in 40 optimized scatterers. Since the model uses diagonal scattering matrices, the number of trainable scattering coefficients is reduced by 50%.

The receiver-side data are generated on two grids. First, a sparse receiver grid consisting of a $30 \times 30$ planar grid (with $x, y \in [A, 4A], z = 0$), is used for model training. Second, a dense validation grid consisting of a $125 \times 125$ planar grid (with $x, y \in [3A, 8A], z = -8\lambda$) is generated for model teat or validation.

To generate the ground-truth field, each scatterer is assigned a diagonal scattering matrix, and the full multi-scatterer coupling is iteratively solved. Particularly, the relaxation factor and the number of iterations of SOR are respectively set to 0.5 and 10. The result evaluated on the dense validation grid is taken as the ground truth. For the equivalent lower-order model, the scattering matrices and the position offsets of the virtual scatterers are jointly optimized by Adam [43] whose number of training epochs is set to 3000.

The training loss is defined as the normalized field error on the sparse receiver grid. After optimization, the learned equivalent model is evaluated again on the dense validation grid to generate the predicted radiomap. Since the direct field is known once the transmit beam is specified, the experiments focus on reconstructing the scattered-field radiomap. The total radiomap can be obtained by adding the direct component to the reconstructed scattered component.

For the choice of matrix inversion solver, Jacobi, Gauss-Seidel, and SOR methods under different numbers of scatterers are investigated. The results in Fig. 6 compares the convergence of a representative scattering coefficient under different iterative solvers and scatterer numbers. In the case of $N_s = 20$, all three methods converge to similar values, while the convergence curve of SOR is smoother. In the case of $N_s = 30$, both Jacobi and Gauss-Seidel diverge in the later iterations, while SOR remains convergent, which complies with our discussion in II-E.

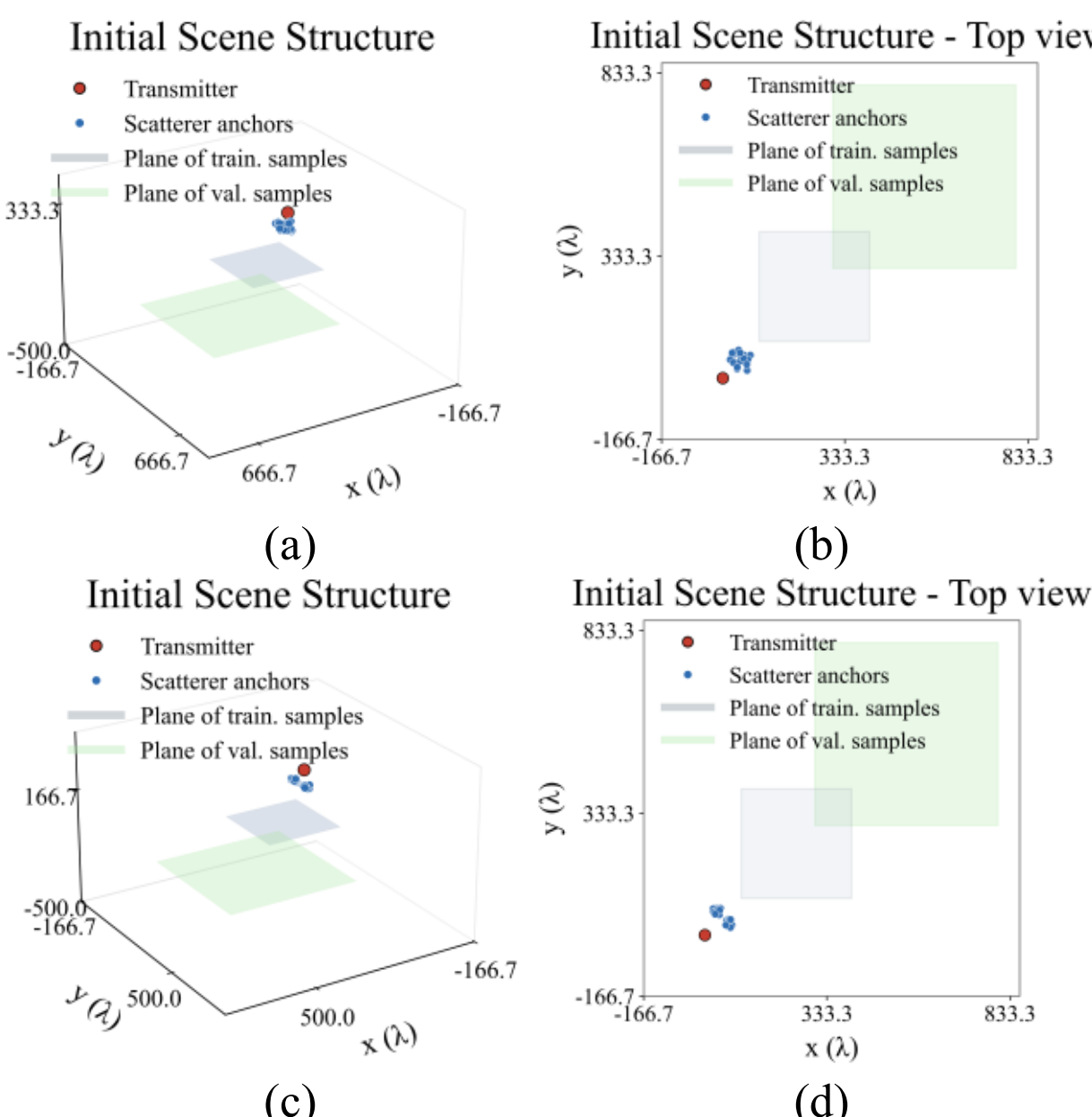

**Fig. 5.** Initial 3D configuration and their top views of the simulation setup (Ns = 20). The subfigure (a) and (b) correspond to the one cluster scenario, whereas (c) and (d) illustrate the two-cluster scenario.

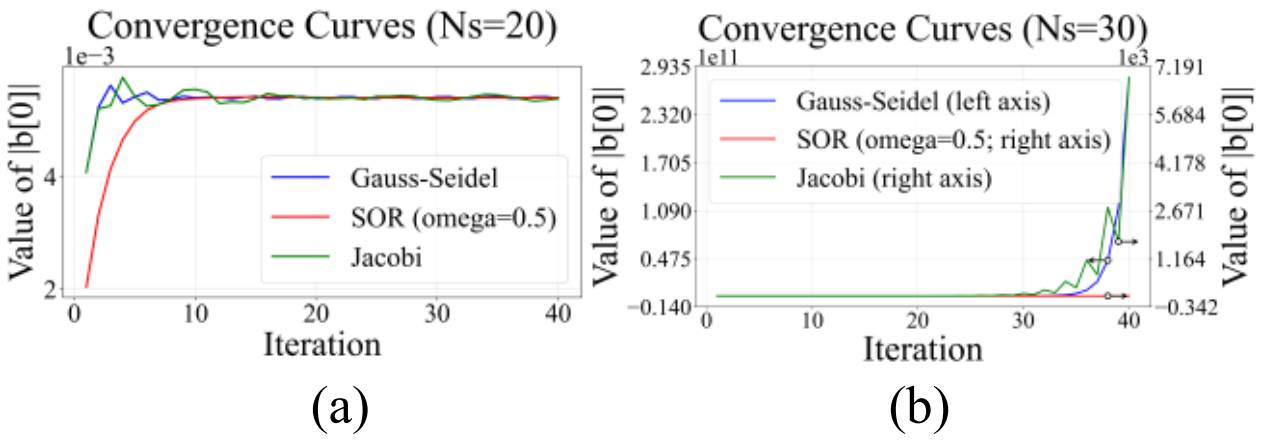

**Fig. 6.** Convergence behavior of the Jacobi, Gauss–Seidel, and SOR iterative methods. (a) Convergence curves for $N_s = 20$; (b) Convergence curves for $N_s = 30$.

### *B. Training for the Equivalent Models*

This subsection is dedicated to verifying that the proposed order-reduction method can produce the same field as the original high-order model. Three models are considered: a high-order target model (namely, target model), an order-preserving equivalent model that uses the same order with the high-order target model (namely, same-order model), and the proposed order-reduced equivalent model (namely, low-order model). In the target model, the scattering matrices and scatterer positions are prescribed, and the generated sparse training samples and dense validation radiomaps are treated as ground truth. The order-preserving equivalent model uses the same number of scatterers and the same multipole order as the target model, serving as a reference for evaluating the proposed low-

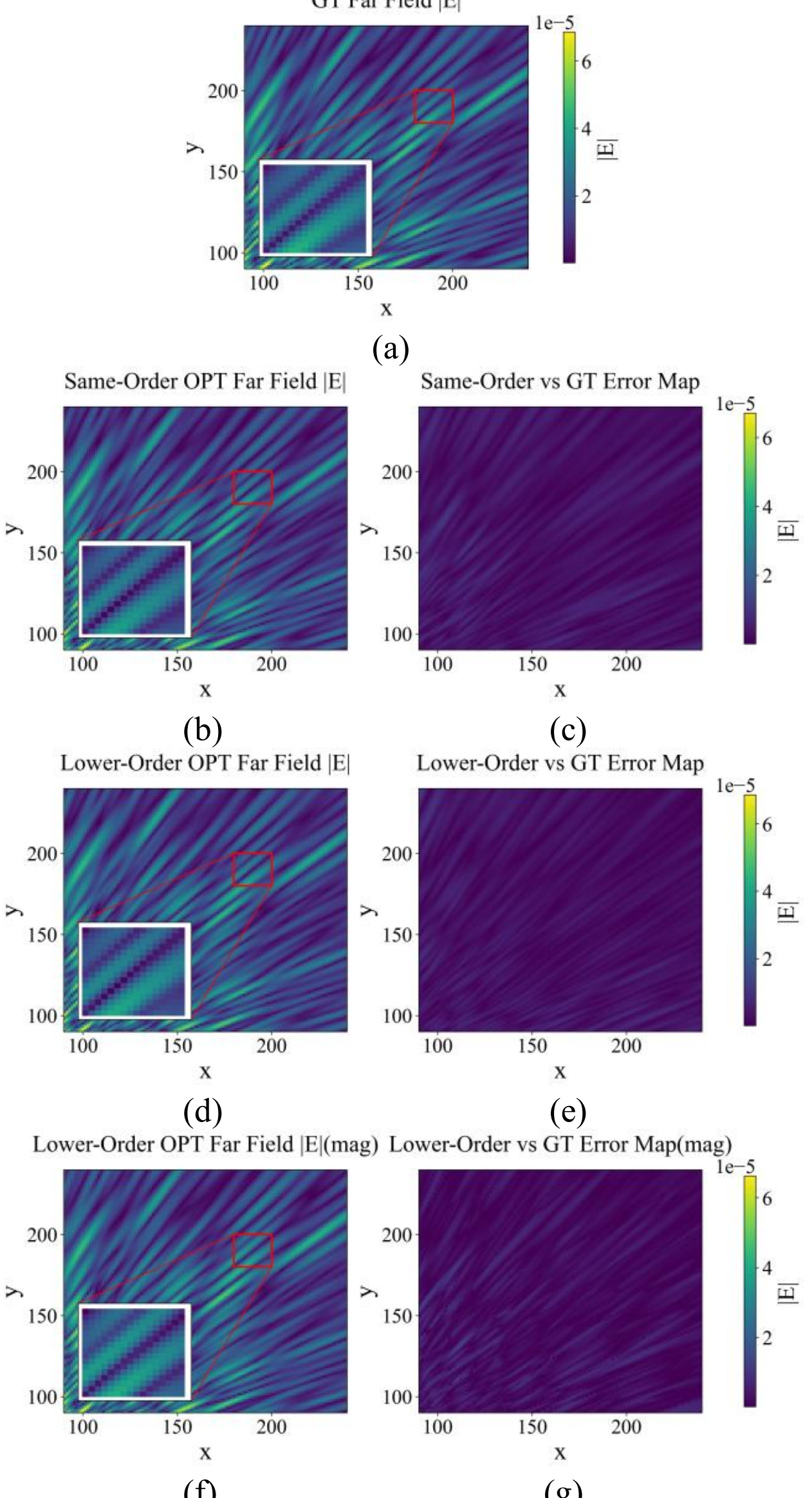

**Fig. 7.** Results of radiomap reconstruction: (a) Far-field ground-truth; (b) Result of the order-preserving model under complex training data; (c) Order-preserving model error map ($| E_{\mathrm{opt}} - E_{\mathrm{GT}} |$) of complex-field results; (d) Result of the order-reduced model under complex training data; (e) Order-reduced model error map ($| E_{\mathrm{opt}} - E_{\mathrm{GT}} |$) of complex-field results; (f) Result of the order-reduced model under magnitude-only training data; (g) Magnitude-only error map of order-reduced model, e.g., $|| E_{\mathrm{opt}} | - | E_{\mathrm{GT}} ||$

order equivalent model. The same receiver grids are used for all three models.

To systematically evaluate the proposed framework, we conduct four complementary sets of comparative simulations. As the common setup, in each training epoch, 15% of the samples from the "training plane grids × beam numbers" training set are randomly selected to form a mini-batch, as illustrated in Fig. 4.

*1) Comparative Study on Model Truncation-Order*

TABLE I
ERROR COMPARISONS FOR TRUNCATION-ORDER AND PHASE-INFORMATION

| Model (data) | Order-preserve (complex) | Order-reduced (complex) | Order-reduced (magnitude) |
|---|---|---|---|
| MSE | 1.44e-11 | 1.21e-11 | 1.05e-11 |
| MAE | 3.37e-06 | 3.09e-06 | 2.50e-06 |
| NMSE | 3.40e-02 | 2.86e-02 | 2.48e-02 |

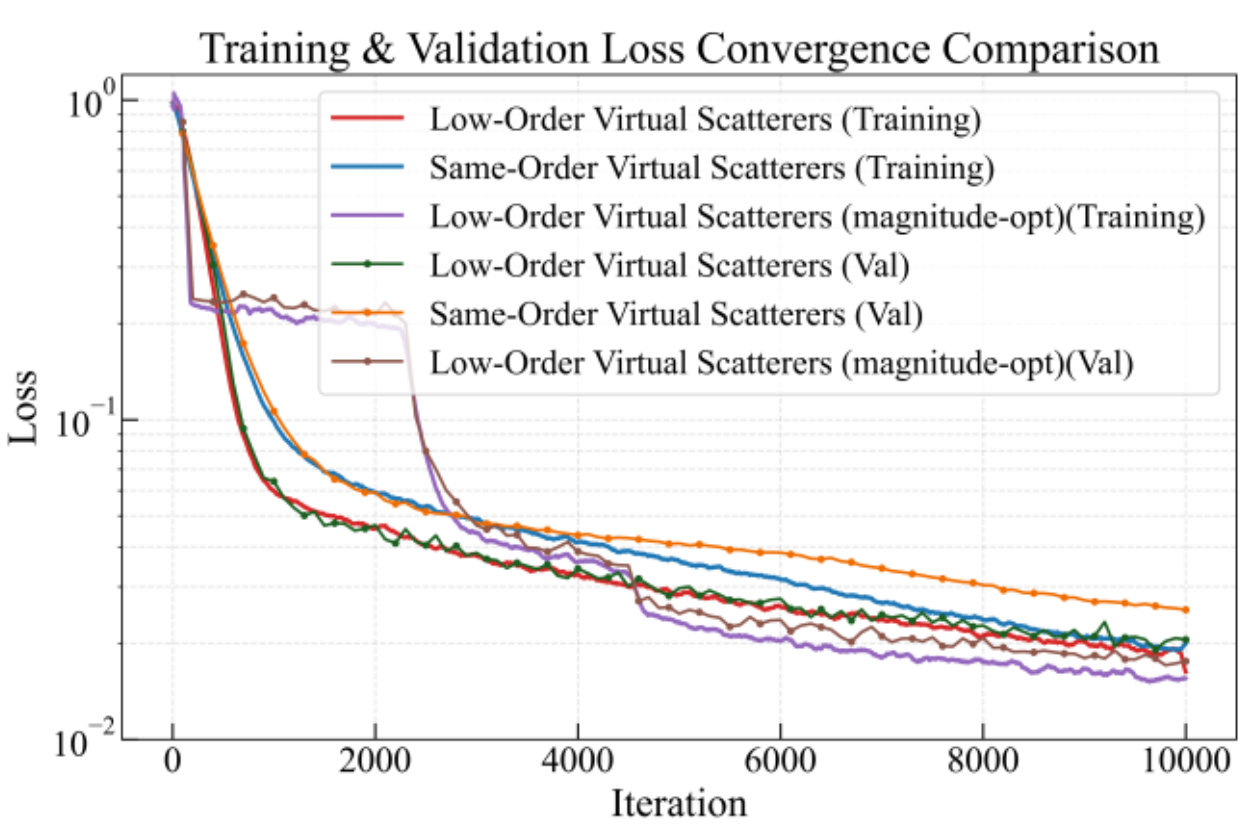

**Fig. 8.** Training and validation loss convergence for the order-preserving equivalent model, order-reduced equivalent model with complex training data, and order-reduced equivalent model with the phaseless CSI data.

First, we compare the order-preserving model with the order-reduced equivalent models to examine the effect of different truncation-orders. The corresponding training objectives are given in (29a) and (32), respectively.

*2) Ablation Study on Phase Information*

Second, using the order-reduced equivalent model, we investigate the effect of phase information in the supervision data. Specifically, we compare training with complex CSI data against training with magnitude-only data. The two settings adopt the same model structure but use different training objectives, as defined in (29a) and (29d), respectively.

Fig. 7 compares the reconstructed far-field radiomaps obtained by the order-preserving model, the order-reduced model with complex training data supervision, and the order-reduced model with magnitude-only supervision. It can be seen that the reconstructed result preserves most of the spatial distribution of the ground-truth field, including subtle localized variations. For quantitative evaluation, the MSE (Mean Squared Error), MAE (Mean Absolute Error) and NMSE (Normalized Mean Squared Error) of respective models are given in Table 1. It is evident that all these three models achieve a satisfactory reconstruction accuracy. Fig. 8 monitors the training and validation loss of respective models. It can be seen that all the training and validation losses decrease to low values, indicating the certain degree of generalization ability.

*3) Comparative Study on Clustering Processing*

Third, to verify the proposed inter-cluster approximation in Section II-F, we consider the two-cluster scene shown in Fig. 5(c). The results presented in Fig. 9 (a), (b), and (d) shows the radiomap generated by the forward problem of the full-coupling model and the inter-cluster model, as well as a clustered order-reduced model optimized in the corresponding inverse problem. The error map between the full-coupling model and inter-

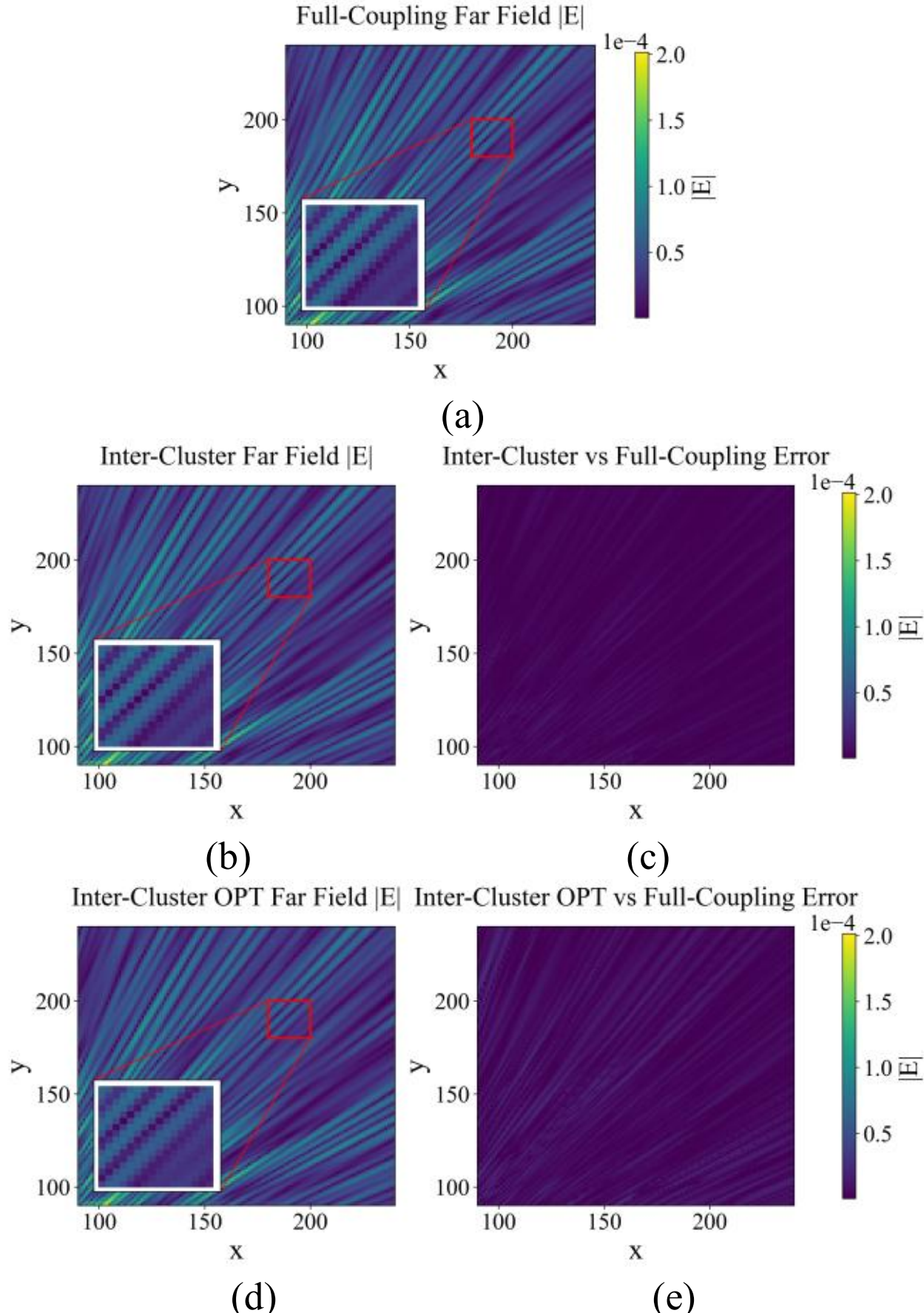

**Fig. 9.** Validation of the proposed inter-cluster approximation. (a) Full-coupling reference field; (b) Forward inter-cluster approximation; (c) Error map between (a) and (b); (d) Inter-cluster optimized radiomap using order-reduced equivalent model. (e) Error map between (a) and (d).

TABLE II

ERROR COMPARISONS FOR CLUSTERING PROCESSING

| Model | Clustered order-preserved (forward) | Culstered order-reduced (optimized) |
|---|---|---|
| MSE | 1.28e-11 | 6.79e-11 |
| MAE | 2.78e-06 | 6.56e-06 |
| NMSE | 4.29e-03 | 2.28e-02 |

cluster order-preserved model is shown in Fig. 9(c), while the error map between the full-coupling model and the clustered order-reduced model is shown in Fig, 9(e). For quantitative evaluation, the MSE (Mean Squared Error), MAE (Mean Absolute Error) and NMSE (Normalized Mean Squared Error) of respective models are given in Table 2.

*4) Ablation Study on Position Optimization*

Last, we investigate the necessity of position optimization merely by comparing the learning curves of two variable settings: optimizing only the scattering matrices (*T-only-opt*) versus jointly optimizing both matrices and spatial offsets (*Joint-opt*). As observed in Fig. 10, *T-only-opt* exhibits higher overall losses and suffers from overfitting after 800 epochs, since its degree-of-freedom (DoF) is insufficient to capture the complex target scattering behavior. Conversely, *Joint-opt* achieves significantly lower final training and validation losses, indicating that spatial offsets provide the crucial additional DoF required for accurate representation.

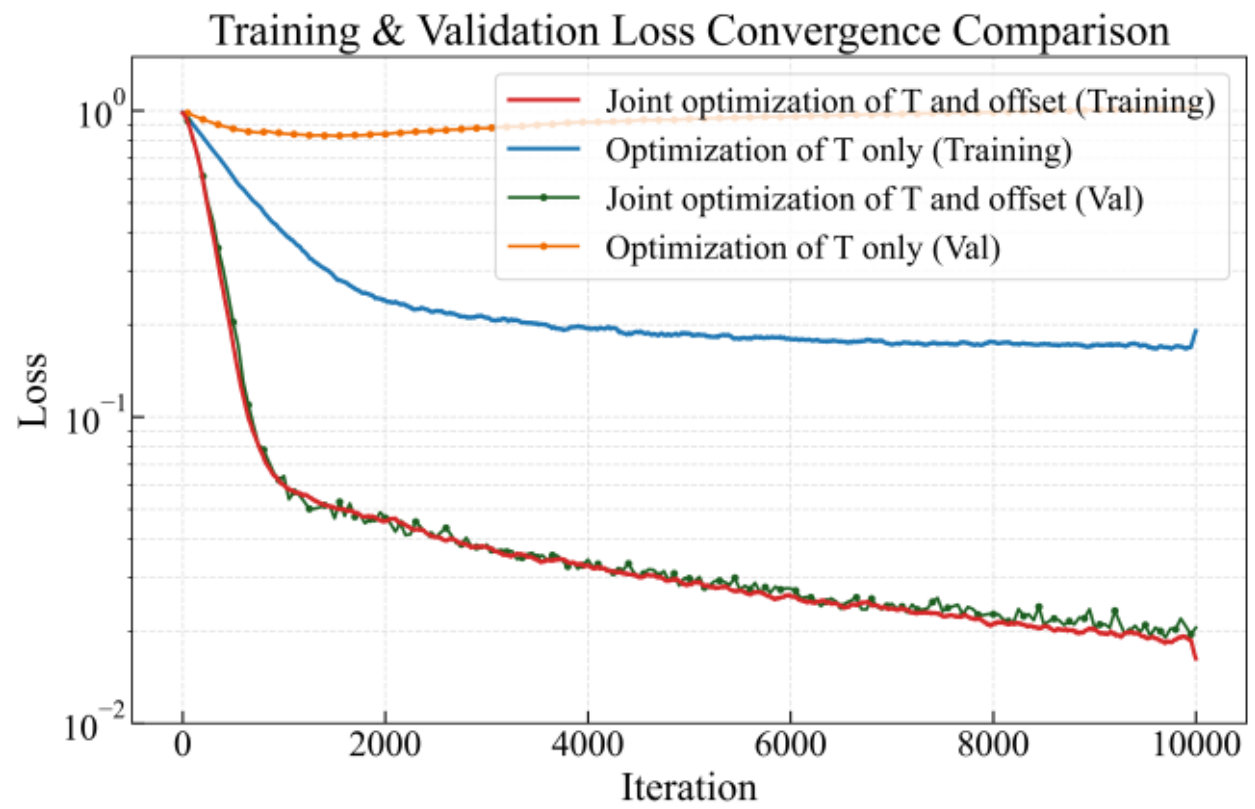

**Fig. 10.** Training and validation loss convergence comparison between two optimization strategies: joint optimization of T and offset, and optimization of T only.

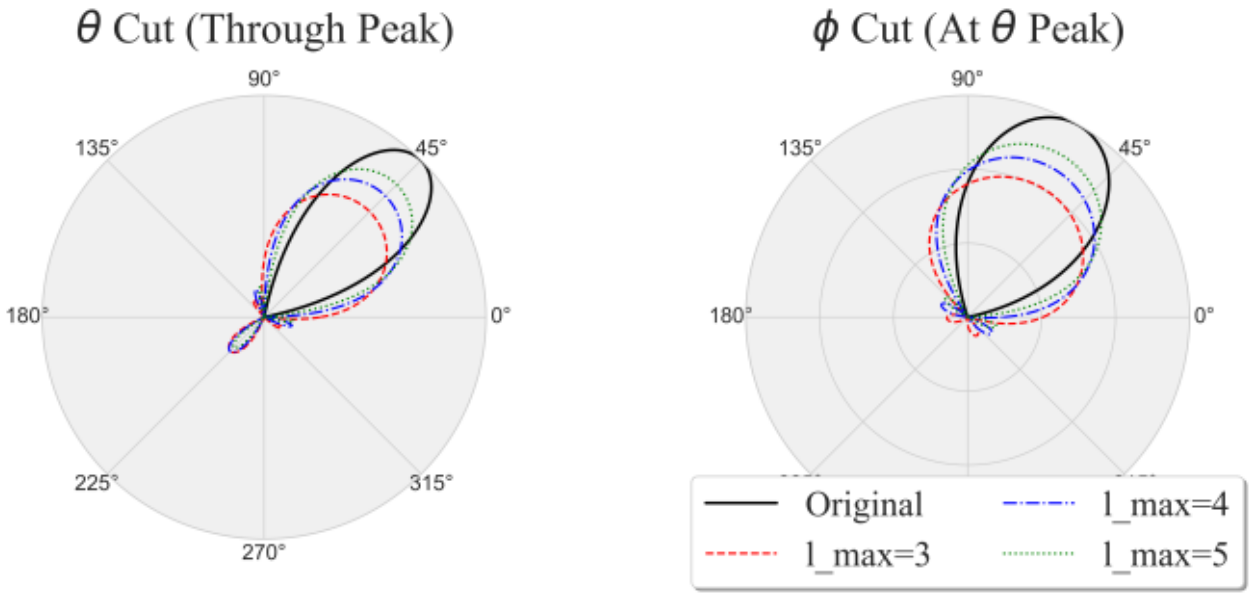

**Fig. 11.** Original beams and fitted beams comparison. Left: $\theta$ cut view; Right: $\varphi$ cut view.

### *C. Beam Synthesis and Radiomap Prediction*

In this subsection, we evaluate the performance of our model on beams that were not seen during training. The beam pattern is designed as

$$F(\theta,\varphi) = A\exp\left(-\frac{k_\theta(\theta-\theta_0)^2 + k_\varphi(\varphi-\varphi_0)^2}{\sigma^2}\right) \qquad (42)$$

where $k_\theta$, $k_\varphi$ and $\sigma$ are parameters controlling the shape of the beam, $\theta_0$ and $\varphi_0$ are steering directions. The beam expansion coefficients in the spherical-harmonic space can be obtained by (34) and (35). Fig. 11 shows one original beam as well as the beam patterns reconstructed using truncated spherical-harmonic expansions. The maximum spherical-harmonic orders are 3, 4 and 5, respectively. It can be seen that the higher the truncation order is, the closer the synthesized beam pattern is to the original beam pattern.

For beam-wise radiomap prediction, the scatterer parameters are trained using 16 directional beams. The centers of these training beams are steered over $\boldsymbol{\theta}_0 = \{90°, 120°, 150°, 180°\}$ and $\boldsymbol{\varphi}_0 = \{0°, 30°, 60°, 90°\}$ (i.e. the total number of training beams is 4 × 4 = 16). The maximum order of the training beam fitting is 3 (i.e., totally $(3+1)^2 = 16$ spherical harmonics are used for fitting). Two additional beams, whose steering directions are not included in the training set, are used for testing. For a testing beam, the ground truth radiomaps are calculated based on the optimized forward model, i.e.

$$\boldsymbol{b}_{test} = \boldsymbol{Q}_{model}\boldsymbol{s}_{test} = (\boldsymbol{I}-\boldsymbol{TH})^{-1}\boldsymbol{TH}_s\boldsymbol{s}_{test}. \qquad (43)$$

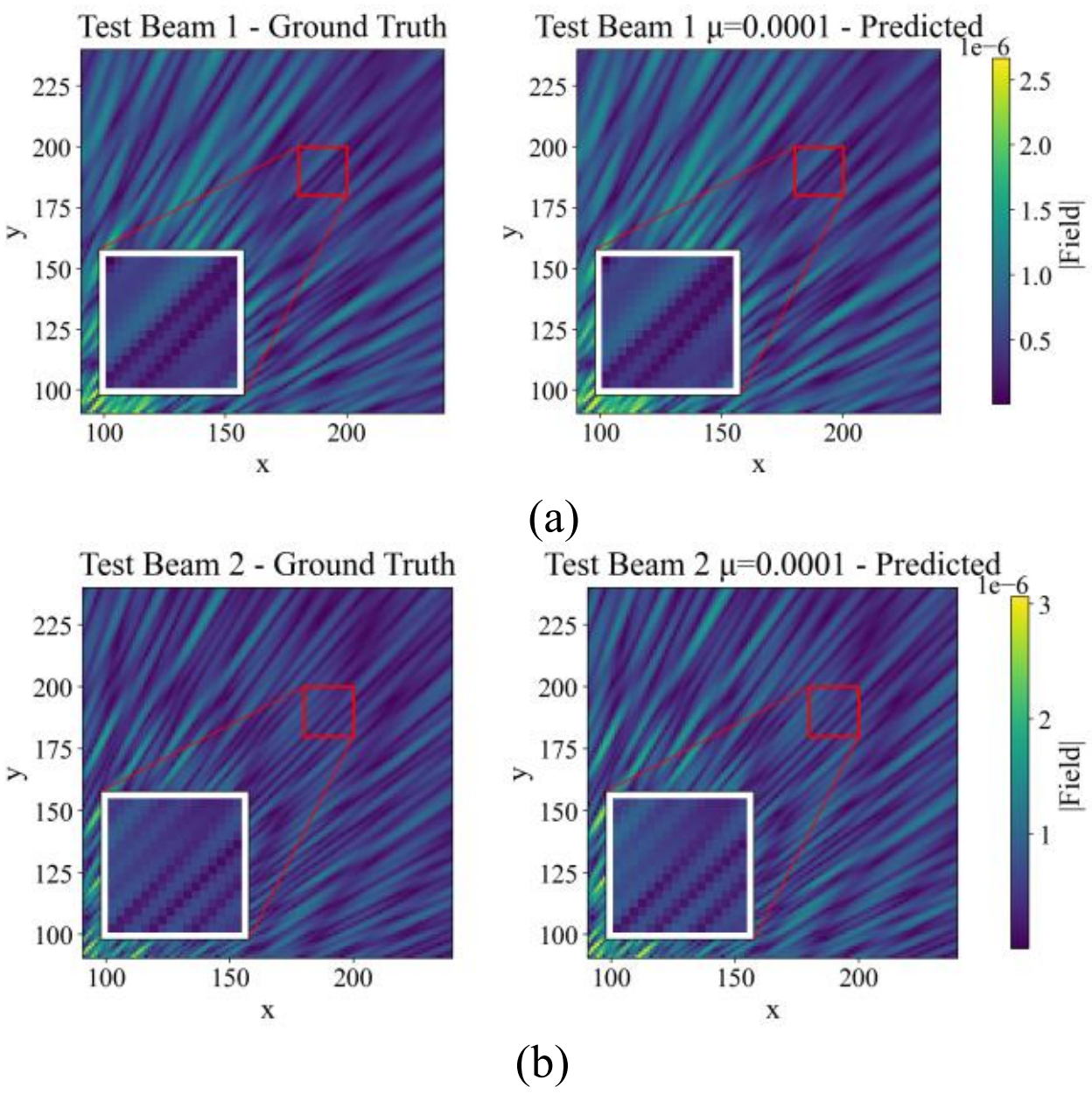

**Fig. 12.** Radiomap comparison under testing beams. (a): Ground truth radiomap and predicted radiomap for test beam 1 ($\theta_0 = 135°, \varphi_0 = 45°$); (b): Test beam 2 ($\theta_0 = 100°, \varphi_0 = 80°$).

where $\boldsymbol{Q}_{model}$ denotes the model-based beam generalization matrix, and $\boldsymbol{s}_{test}$ is obtained by projecting the testing beam pattern onto the spherical-harmonic basis.

For prediction, the data-driven beam generalization matrix $\boldsymbol{Q}_{data}$ is estimated from the training beams by (39). Since the inversion associated with $\boldsymbol{S}$ can be ill-conditioned, a regularized pseudo-inverse is used to obtain the scattering coefficients:

$$\boldsymbol{b}_{new} = \boldsymbol{Q}_{data}\boldsymbol{s}_{test} = \boldsymbol{B}\boldsymbol{S}^{\boldsymbol{H}}(\boldsymbol{S}\boldsymbol{S}^{\boldsymbol{H}} + \mu\boldsymbol{I})^{-1}\boldsymbol{s}_{test} \quad (44)$$

In our simulations, $\mu$ = 1e-4 is chosen to balance the numerical stability and the reconstruction fidelity. The scattering radiomap distribution is computed from the scattering coefficients $\boldsymbol{b}$ of scatterers, as per (16) and (19).

Fig. 12 shows that the predicted radiomaps closely match the ground-truth, confirming the beam-wise generalization capability of the proposed model.

### *D. CFR Interpolation*

In this subsection, we evaluate the CFR interpolation capability of our learning-based, physics-embedded multi-scatterer model. The order-reduced equivalent model is optimized at only five sparse anchor frequencies, uniformly distributed from 0.99 to 1.01 (normalized frequency). During optimization, the equivalent scatterer positions ($\boldsymbol{R}_{eq}$) are shared across all frequencies, while the frequency-dependent T-matrix ($\boldsymbol{T}_{\nu}$) is optimized independently at each anchor. Intermediate frequencies are reserved for testing.

To evaluate the advantage of the physics-embedded representation, we compare two interpolation strategies. The first is a direct CFR interpolation, which separately interpolates the real and imaginary parts of the channel response across the anchor frequencies. The second is our T-matrix interpolation strategy, which first interpolates the learned $\boldsymbol{T}_{\nu}$ matrix across the frequency domain and subsequently evaluates the physics-based forward model to synthesize the CFR.

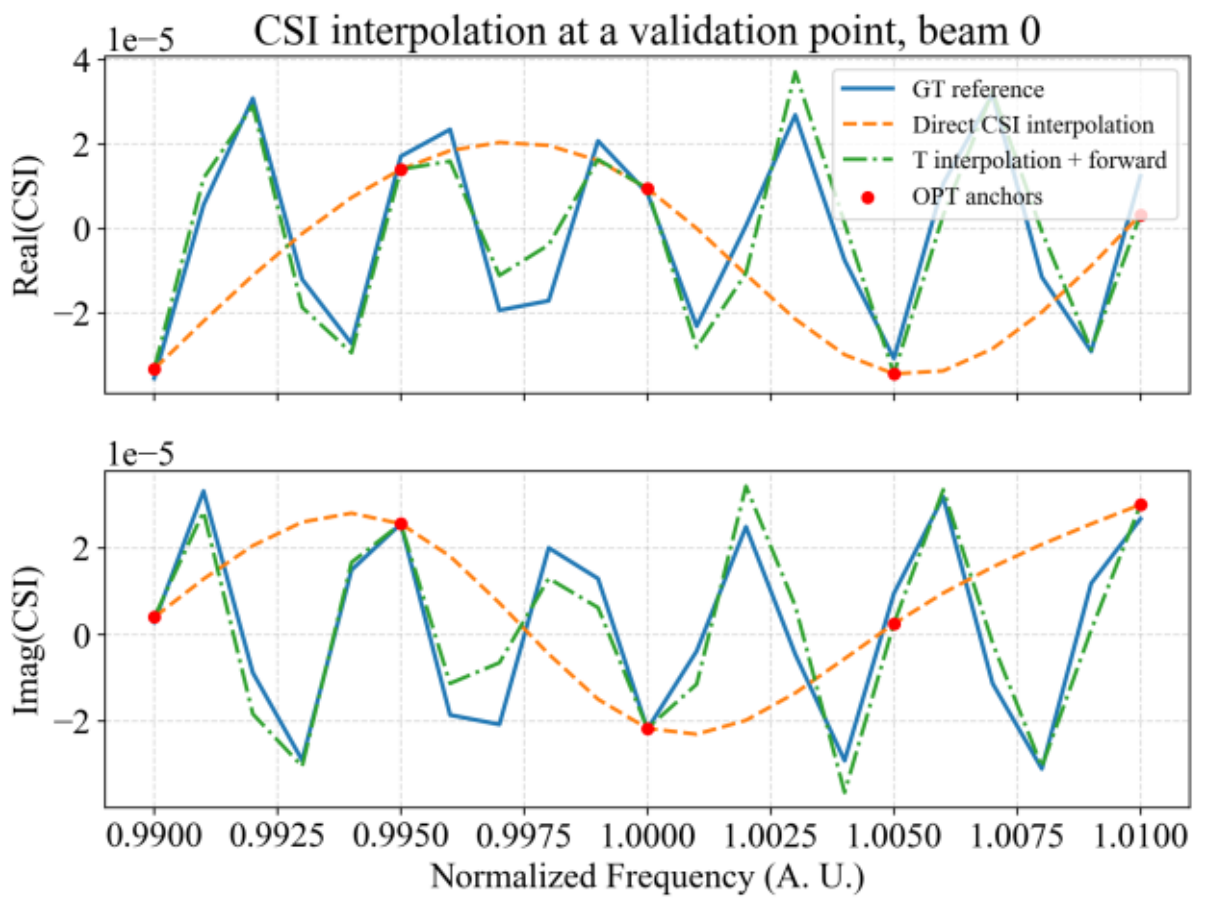

**Fig. 13.** Frequency-domain CSI interpolation over 21 normalized frequency samples within a narrowband interval at a representative validation point for beam 0.

As visualized in Fig. 13 for a representative validation receiving point, direct CFR interpolation fails to capture the rapid, frequency-dependent fluctuations of the complex channel response. Conversely, interpolating the T-matrix and applying the forward model accurately tracks the ground truth trajectory. This improvement arises because the CFR is an end-to-end channel quantity that includes rapidly varying propagation phases, whereas the learned T-matrices represent local scatterer responses that change more smoothly over the considered bandwidth. Therefore, our model preserves the frequency-dependent complex channel trajectory.

## V. Conclusion

In this work, we construct a multi-scatterer channel model from the wave perspective. In this model, the source radiation and scatterer response are expanded as superpositions of spherical-wave modes, capturing the multi-path effect. Both the forward problem and its corresponding inverse optimization problem are formulated. An approximate model is introduced, where a large number of denser scatterers are used with fewer and simpler expansion modes. Simulation results show the accuracy of radiomap reconstructed from sparse field samples. Our model can be well generalized to spatial locations, beam configurations, and frequency points that are unseen in the training set.